\documentclass[longauth]{aa} 
\usepackage{amsmath}	
\usepackage{amssymb}	
\makeatletter
\renewcommand\@biblabel[1]{}

\makeatother
\usepackage{graphicx}
\usepackage{txfonts}
\begin{document}

   \title{On the nature of a shell of young stars in the outskirts \\ of the Small Magellanic Cloud}


\author{David Mart\'\i nez-Delgado\inst{1},
A.\ Katherina Vivas$^{2}$, Eva K.\ Grebel$^{1}$, Carme Gallart$^{3,4}$, Adriano Pieres$^{5,6}$, Cameron P.~M.~Bell$^{7}$, Paul Zivick$^{8}$, Bertrand Lemasle$^{1}$, L. Clifton Johnson$^{9}$, Julio A. Carballo-Bello$^{10,11}$, Noelia E. D. No\"el$^{12}$, Maria-Rosa L. Cioni$^{7}$, Yumi Choi$^{13, 14}$, Gurtina Besla$^{13}$, Judy Schmidt$^{15}$, Dennis Zaritsky$^{13}$, Robert A.\ Gruendl$^{16,17}$, Mark Seibert$^{18}$, David Nidever$^{19}$, Laura Monteagudo$^{3,4}$, Mateo Monelli$^{3,4}$, Bernhard Hubl$^{20}$, Roeland van der Marel$^{21, 22}$, Fernando J. Ballesteros$^{23}$, Guy Stringfellow$^{24}$, Alistair Walker$^{2}$, Robert Blum$^{19}$, Eric F. Bell$^{25}$, Blair C. Conn$^{26}$, Knut Olsen$^{19}$, Nicolas Martin$^{27,28}$, 
 You-Hua Chu$^{29,30}$, Laura Inno$^{28}$, Thomas J. L. Boer$^{12}$, Nitya Kallivayalil$^{8}$, Michele De Leo$^{12}$, Yuri Beletsky$^{31}$, Ricardo R. Mu\~noz $^{32}$}

\institute{$^{1}$Astronomisches Rechen-Institut, Zentrum f\"ur Astronomie der
Universit\"at Heidelberg, M\"onchhofstr.\ 12--14, 69120 Heidelberg,
Germany\\
$^{2}$Cerro Tololo Inter-American Observatory, National Optical Astronomy Observatory, Casilla 603, La Serena, Chile\\
 $^{3}$Instituto de Astrof\'isica de Canarias (IAC), Calle V\'ia L\'actea s/n, E-38205 La Laguna, Tenerife; Spain\\
$^{4}$Departamento de Astrof\'isica, Universidad de La Laguna (ULL), E-38206 La Laguna, Tenerife; Spain\\ 
$^{5}$ Laborat\'orio Interinstitucional de e-Astronomia - LIneA, Rua Gal. Jos\'e Cristino 77, Rio de Janeiro, RJ - 20921-400, Brazil \\
$^{6}$ Observat\'orio Nacional, Rua Gal. Jos\'e Cristino 77, Rio de Janeiro, RJ - 20921-400, Brazil \\
$^{7}$ Leibniz-Institut f\"ur Astrophysik Potsdam, An der Sternwarte 16, D-1442 Potsdam, Germany \\
$^{8}$ Department of Astronomy, University of Virginia, 530 McCormick Road, Charlottesville, VA 22904, USA \\
$^{9}$ CIERA and Department of Physics and Astronomy, Northwestern University, 2145 Sheridan Road, Evanston, IL 60208, USA\\ 
$^{10}$ Instituto de Astrofisica, Pontificia Universidad Catolica de Chile, Av. Vicuna Mackenna 4860, Macul 7820436, Santiago, Chile \\
$^{11}$Chinese Academy of Sciences South America  Center for Astronomy, National Astronomical Observatories, CAS, Beijing 100101, China \\ 
$^{12}$ Department of Physics, University of Surrey, Guildford GU2 7XH, UK\\
$^{13}$ Steward Observatory, University of Arizona, 933 North Cherry Avenue,Tucson, AZ 85721\\
$^{14}$ Department of Physics, Montana State University, P.O. Box 173840, Bozeman, MT 59717-3840, USA\\
$^{15}$ Astrophysics Source Code Library, University of Maryland, 4254 Stadium Drive College Park, MD 20742, USA\\
$^{16}$ National Center for Supercomputing Applications, 1205 West Clark St., Urbana, IL 61801, USA
$^{17}$ Department of Astronomy, University of Illinois, 1002 W. Green Street, Urbana, IL 61801, USA 
$^{18}$ The Observatories of the Carnegie Institution for Science, 813 Santa Barbara St., Pasadena, CA 91101\\
$^{19}$ National Optical Astronomy Observatory, 950 N. Cherry Ave, Tucson, AZ 85719, USA\\
$^{20}$ Astrophoton Observatory, Edtfleck 12, A-4542 Nussbach, Austria\\
$^{21}$ Space Telescope Science Institute, 3700 San Martin Drive, Baltimore,  MD 21030\\
$^{22}$ Center for Astrophysical Sciences, Department of Physics \& Astronomy, Johns Hopkins University, Baltimore, MD 21218, USA\\
$^{23}$ Observatori Astron\`omic, Universitat de Val\`encia, Paterna (Val\`encia), Spain\\
$^{24}$ Center for Astrophysics and Space Astronomy, Department of Astrophysical and Planetary Sciences,University of Colorado 389 UCB Boulder Colorado  80309-0389, USA \\
$^{25}$ Department of Astronomy, University of Michigan, 1085  S. University Ave., Ann Arbor, MI 48109 USA\\
$^{26}$ Research School of Astronomy and Astrophysics, Australian National University, Canberra, ACT 2611, Australia\\
$^{27}$ Universit\'e de Strasbourg, CNRS, Observatoire astronomique de Strasbourg, UMR 7550, F-67000 Strasbourg, France\\
$^{28}$ Max-Planck-Institut f\"{u}r Astronomie, K\"{o}nigstuhl 17, D-69117 Heidelberg, Germany \\
$^{29}$Institute of Astronomy and Astrophysics, Academia Sinica, P.O. Box 23-141, Taipei 10617, Taiwan, R.O.C.\\
$^{30}$Department of Astronomy, University of Illinois at Urbana-Champaign, 1002 West Green Street, Urbana,IL 61801, USA\\
$^{31}$ Las Campanas Observatory, Carnegie Institution of Washington, La Serena, Chile\\
$^{32}$ Departamento de Astronom\'ia, Universidad de Chile, Camino del Observatorio 1515, Las Condes, Santiago, Chile}


   \date{}

 
  \abstract
   {Understanding the evolutionary history of the Magellanic Clouds requires an in-depth exploration and characterization of the stellar content in their outer regions, which ultimately are key to tracing the epochs and nature of past interactions.}
   {We present new deep images of a shell-like over-density of stars in the outskirts of the Small Magellanic Cloud (SMC). The shell, also detected in photographic plates dating back to the fifties, is located at $\sim 1.9\degr$ from the center of the SMC in the north-east direction.}
   {The structure and stellar content of this feature were studied with multi-band, optical data from the Survey of the MAgellanic Stellar History (SMASH) carried out with the Dark Energy Camera on the Blanco Telescope at Cerro Tololo Inter-American Observatory. We also investigate the kinematic of the stars in the shell using the {\it Gaia} Data Release 2.}
   {The shell is composed of a young population with an age $\sim 150$ Myr, with no contribution from an old population. Thus, it is hard to explain its origin as the remnant of a tidally disrupted stellar system. The spatial distribution of the young main-sequence stars shows a rich sub-structure, with a spiral arm-like feature emanating from the main shell and a separated small arc of young stars close to the globular cluster NGC 362. We find that the absolute $g$-band magnitude of the shell is M$_{g,shell} = -10.78\pm 0.02$, with a surface brightness of $\mu_{g,shell} = 25.81\pm 0.01$ mag~arcsec$^{-2}$}
   {We have not found any evidence that this feature is of tidal origin or a bright part of a spiral arm-like structure. Instead, we suggest that the shell formed in a recent star formation event, likely triggered by an interaction with the Large Magellanic Cloud and/or the Milky Way, $\sim$ 150 Myr ago.}

   \keywords{Galaxies:individual:Small Magellanic Cloud -- Local Group --Magellanic Clouds -- Galaxies:photometry --Galaxies:structure} 
 \authorrunning{Mart\'\i nez-Delgado et al.}
 
  \maketitle
%

\section{Introduction}

The Magellanic Clouds are the largest satellites of the Milky Way (MW), and
the only irregular galaxies in its immediate surroundings.  The
existence of one or even two such gas-rich, massive satellites close
to a Milky-Way-sized halo has been shown to be quite rare (e.g., Busha
et al. 2011; Gonz\'alez et al.\ 2013; Rodr{\'{\i}}guez-Puebla {\it et al.}\
2013; Boylan-Kolchin {\it et al.} 2011; Patel, Besla \& Sohn 2017). There is strong evidence from the motions and distributions of the stellar and gaseous constituents of the Magellanic Clouds that interactions between them have occurred in the past. High-precision proper motion measurements with the Hubble Space Telescope revealed that, contrary to earlier belief, the Magellanic Clouds are probably just completing their first passage around the Milky Way (Besla {\it et al.} 2007, 2010; Kallivayalil {\it et al.} 2006a, 2006b, 2013; Piatek {\it et al.} 2008; Bekki 2011).

 Clearly, the past
interactions between the Clouds have left their marks. One of the most obvious manifestations 
of this interaction is the gaseous Magellanic
Stream (Mathewson et al. 1974), whose trailing and leading arms have
since been traced over more than $200\degr$ across the sky (Putman et
al.\ 1998; Nidever et al.\ 2010).  Both tidal and ram pressure
stripping origins have been suggested for the Stream (see D'Onghia \&
Fox 2016 for a review).  The mass of ionized and atomic Magellanic gas
found outside of the Clouds exceeds the remaining H\,{\sc i} mass in
both Clouds combined (Fox et al.\ 2014) and is of a similar order of
magnitude as the {\it total} mass of the Small Magellanic Cloud (SMC).

\begin{figure*}
	\includegraphics[width=1.0\textwidth]{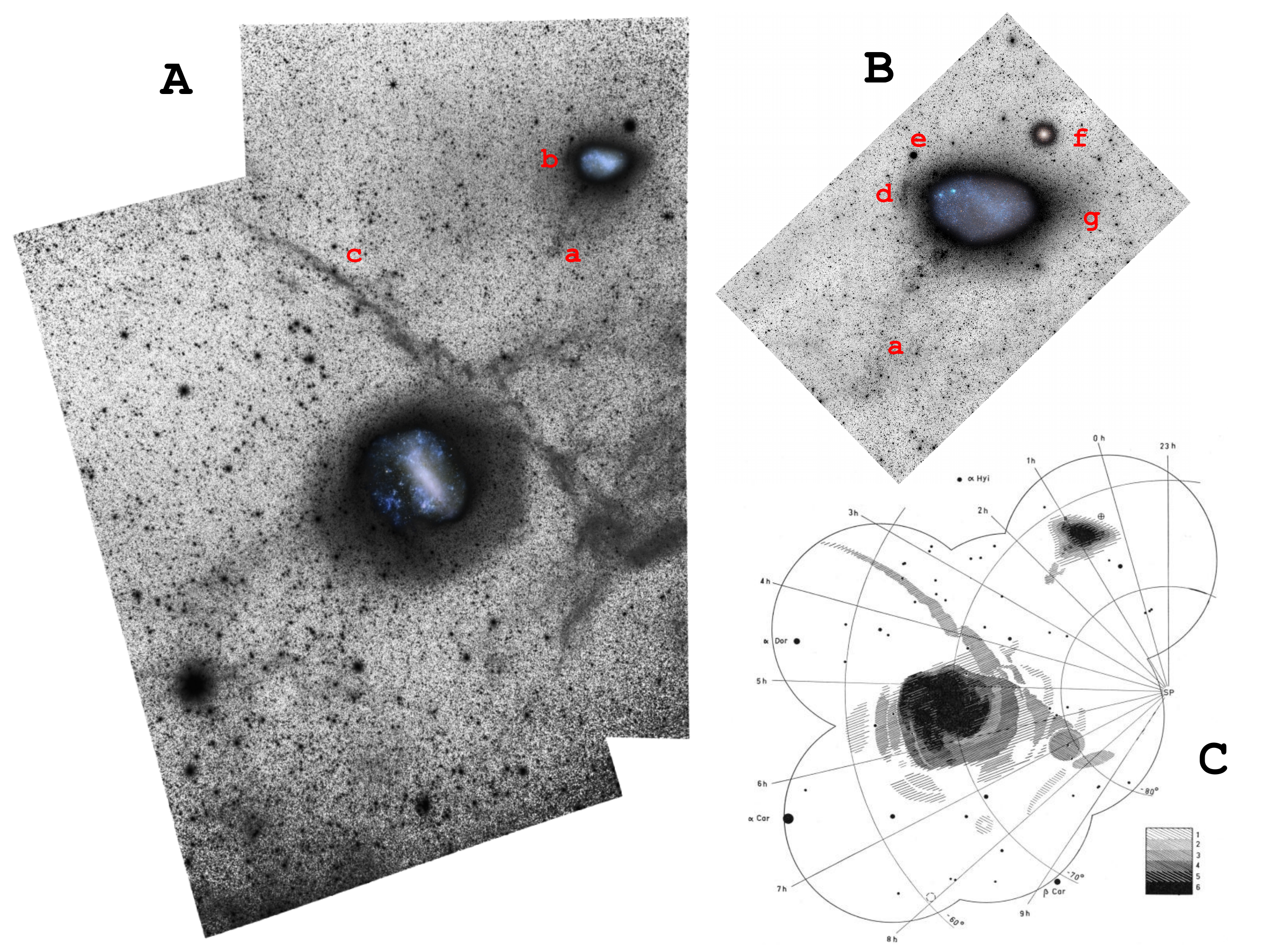}
           \caption{The de Vaucouleurs' photographic study of the Magellanic Clouds revisited with CCD images obtained with Canon telephoto lens from ESO La Silla Observatory. {(\it Panel A):} Full wide-field mosaic of the Clouds obtained with a Canon 50 lens as described in Besla et al. (2016). The  main features described throughout this paper are marked with letters: The Wing (label {\it a}), the outer arm B (label {\it b}) and a well-known foreground Galactic cirrus crossing the field (label {\it c});  {(\it Panel B)}: A zoomed CCD image of the SMC obtained in the same run with a Canon 200 lens showing the shell-like feature (label d) in the outer arm, the SMC south-west plume (label g) and the two globular clusters NGC 362 (label e) and 47 Tuc (label f); {(\it Panel C)}: For a comparison, the sketch based on the 1950s photographic plate material of the Clouds from de Vaucouleurs \& Freeman (1972) already showed all the features detected in our modern deep CCD imaging. (©AAS. Reproduced with permission)} 
    \label{fig-panel}
\end{figure*}

Another prominent feature is the gaseous Magellanic Bridge, which
connects the two Clouds. The Bridge also contains stars, specifically
an irregularly distributed young stellar population with ages of up to
several hundred Myr (e.g., Irwin et al.\ 1985; Skowron et al.\ 2014),
associations, and star clusters (e.g., Bica et al.\ 2008).  Unequivocal evidence of intermediate-age stars was also found in the Magellanic Bridge area  (No\"el et al. 2013; No\"el et al. 2015)  as well as 
unambiguous evidence of a tidal stripping scenario (Carrera et al. 2017). The  density distribution of these inter-cloud stars suggests that they belong to the extended outer
regions of the two Clouds (e.g., Skowron et al.\ 2014;
Jacyszyn-Dobrzeniecka et al. 2017; Wagner-Kaiser \& Sarajedini 2017). This would support a scenario where the Bridge formed in a close encounter between the Clouds some 250 Myr ago (D'Onghia \& Fox 2016 and references therein). On the other hand, Belokurov et al. (2017) report the discovery of a separate, off-set old stellar bridge between the Clouds.
However, Jacyszyn-Dobrzeniecka et al. (2019) found the existence of this old bridge as controversial, based on the analysis of the distribution of RR Lyrae stars in the OGLE data. 

Moreover, the outer regions of both Clouds show distortions, clumps,
arcs, and related overdensities, both in the direction of the Bridge
and elsewhere (e.g., Nidever et al.\ 2013; Casetti-Dinescu et al.\
2014; Besla et al.\ 2016; Belokurov \& Koposov 2016; Mackey et al.\
2016; Belokurov et al.\ 2017; Pieres et al.\ 2017; Subramanian et al.\ 2017; Carrera  et al.\ 2017; Mackey et al.\ 2018; Choi et al. 2018a; Choi et al. 2018b). The last close
encounter between the two Clouds some 100 to 300 Myr ago is not only consistent with the young ages in the
Bridge, but also with a peak in the age distribution of young clusters
in both Clouds (Glatt et al.\ 2010) and with the bimodal ages of Cepheids (Ripepi et al.\ 2017).

The SMC has been particularly affected by past interactions and shows
a highly distorted, amorphous structure in its H\,{\sc i} distribution
(e.g., Stanimirovic et al.\ 2004) and in its younger populations,
while its old populations are symmetrically and regularly distributed
(e.g. Cioni, Habing \& Israel 2000; Zaritsky et al.\ 2000; Haschke et al.\ 2012; Jacyszyn-Dobrzeniecka et
al. 2017; Muraveva et al.  2018).  The SMC has a large line-of-sight depth (e.g., Caldwell \& Coulson 1986; Mathewson et al.\ 1988;
Crowl et al.\ 2001; Subramanian \& Subramaniam 2012; Haschke et al.\
2012; Nidever et al. 2013) and its old stellar population is very extended (No\"el \& Gallart 2007; Nidever et al. 2011).  
 From their study of the 3-D structure
of the SMC  using Cepheid stars, Ripepi et al.\ (2017) found 25-30 kpc (see also Scowcroft et al.\ 2016;
Jacyszyn-Dobrzeniecka et al.\ 2016).  Apart from repeated disruptive encounters between
the Clouds, it has been suggested that the complex structure of the SMC may also
be due to a dwarf-dwarf merger in the distant past
(Bekki \& Chiba 2008).  

\begin{figure*}
\centering
		\includegraphics[width=0.80\textwidth]{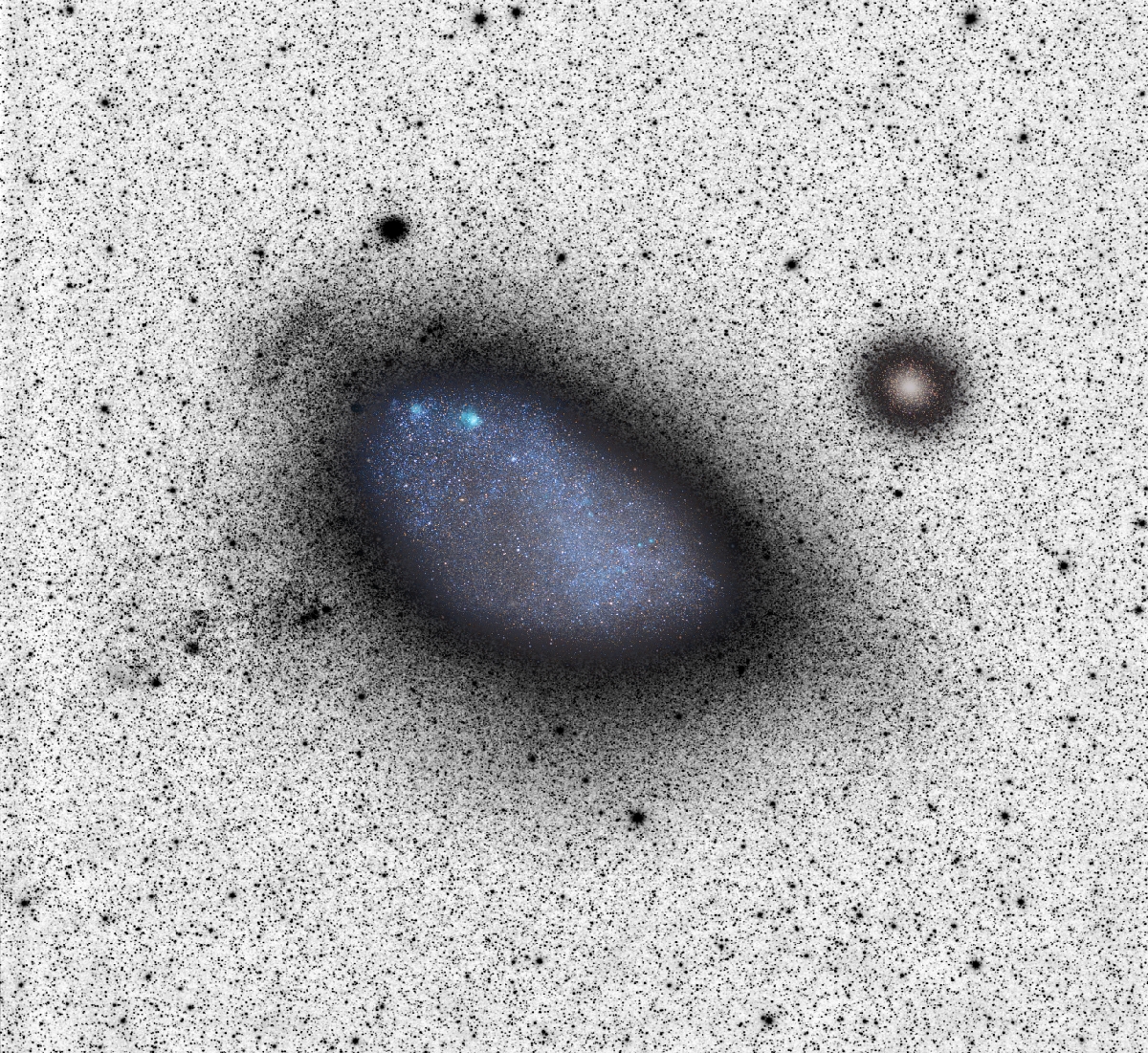}
    \caption{Image of the Small Magellanic Cloud obtained with a Canon
EF 200 mm f/2.8L lens attached to a Canon E0S6D (ISO1600) camera on
2015 October 10 from Hacienda Las Condes (Chile). For illustrative
purposes, a color inset of the disk of the galaxy was superposed on
the linearly stretched, negative image. The total field of view is
$10.2\degr \times 6.8 \degr$, with a pixel scale of
6.7 arcsec~pixel$^{-1}$.}
    \label{fig-canon}
\end{figure*}

  One important approach towards understanding the evolutionary history of the Magellanic Clouds is through deep, multi-colour mapping of
the Clouds, especially of their neglected outskirts, which can still contain clues about the times and nature of past interactions.
As part of our deep, wide-field imaging survey of faint tidal structures around the Magellanic Clouds and other Milky Way satellites
(Besla et al. 2016; Mart\'\i nez-Delgado et al. in preparation) using telephoto lenses, we have detected a coherent shell-like over-density feature embedded in the diffuse "outer arm B" (see de Vaucouleurs \& Freeman 1972) and previously visible in photographic studies of the Clouds dating back to the  1950s (e.g. see Figure 12b in de Vaucouleurs \& Freeman 1972 and Fig. \ref{fig-panel} in this work). Albers et al. (1987) also found evidence of shell-type structures on the north-east side of the SMC based on the analysis of star counts from scanned photographic plates, suggesting they could be associated to faint, spiral arm structure in this region of the SMC. The shell-like feature was also previously noted as ``a distinct linear feature perpendicular to the main elongation of the SMC'' in a stellar density map using
only upper main-sequence stars selected from the Magellanic Cloud Photometric Survey in Zaritsky et al.\ (2000, their Fig. 2). An early color-magnitude diagram of the north-east outer regions of the SMC by Brueck \& Marsoglu (1978) concluded the presence of a young population ($\sim$ 60 Myr) and the association of three star clusters in the {\it outer arm B}, that were interpreted as the evidence of a recent burst of star formation in this SMC region (Bruck 1980).  In this
paper, we study the structure, stellar and gas content, kinematics and possible formation scenarios of this over-density by means of its resolved stellar populations as traced by the imaging data obtained in the Survey of the MAgellanic Stellar History (SMASH, Nidever et al. 2017) and the proper motions available from the {\it Gaia} Data Release 2 (DR2) (Gaia Collaboration et al. 2018).

\section{The Data}

\subsection{Canon 200 lens data}

In our data, the SMC over-density was first detected in deep images
taken with an equipment consisting of a Canon EF 200 mm f/2.8L
II USM and a SBIG STL-11000M CCD camera during two observing runs in
2009 August and September at the European Southern Observatory (La
Silla, Chile).  This setup provided a total field of view (FoV) of
$10\degr \times 7\degr$ and a pixel scale of 9.27 arcsec~pixel$^{-1}$.
A set of 58 individual exposures of 300 sec were taken  in a Luminance
filter (see, e.g., Fig.\ 1 in  Mart\'\i nez-Delgado et al. 2015), with a
total exposure time of 290 minutes. This first image is showed in Fig. 1 ({\it panel B}), 
with the SMC over-density marked with label {\it d}.

Standard data reduction procedures for bias subtraction and flat
fielding were carried out using the CCDRED package in the Image
Reduction and Analysis Facility (IRAF\footnote{IRAF is distributed by
the National Optical Astronomy Observatories, which are operated by
the Association of Universities for Research in Astronomy, Inc., under
cooperative agreement with the National Science Foundation.}). A
detailed analysis of these observations is presented in
Mart\'\i nez-Delgado et al. (in preparation).

To ensure that this shell-like feature was not an artifact or a reflection, a
confirmation wide-field image of the SMC was taken with a different
Canon EF 200 mm f/2.8L lens attached to a Canon E0S6D (ISO1600) camera
in 2015 October 10 from Hacienda Los Andes (Chile).  Figure ~\ref{fig-canon} shows
the resulting image, with a total exposure time of 156 minutes
(obtained by combining 78 individual images with an exposure time of
two minutes). Astrometry of
this image was obtained using the astrometric calibration service 
\url{Astrometry.net} (Lang et al. 2010).

\subsection{SMASH data}

\begin{figure*}
	\includegraphics[width=0.50\textwidth]{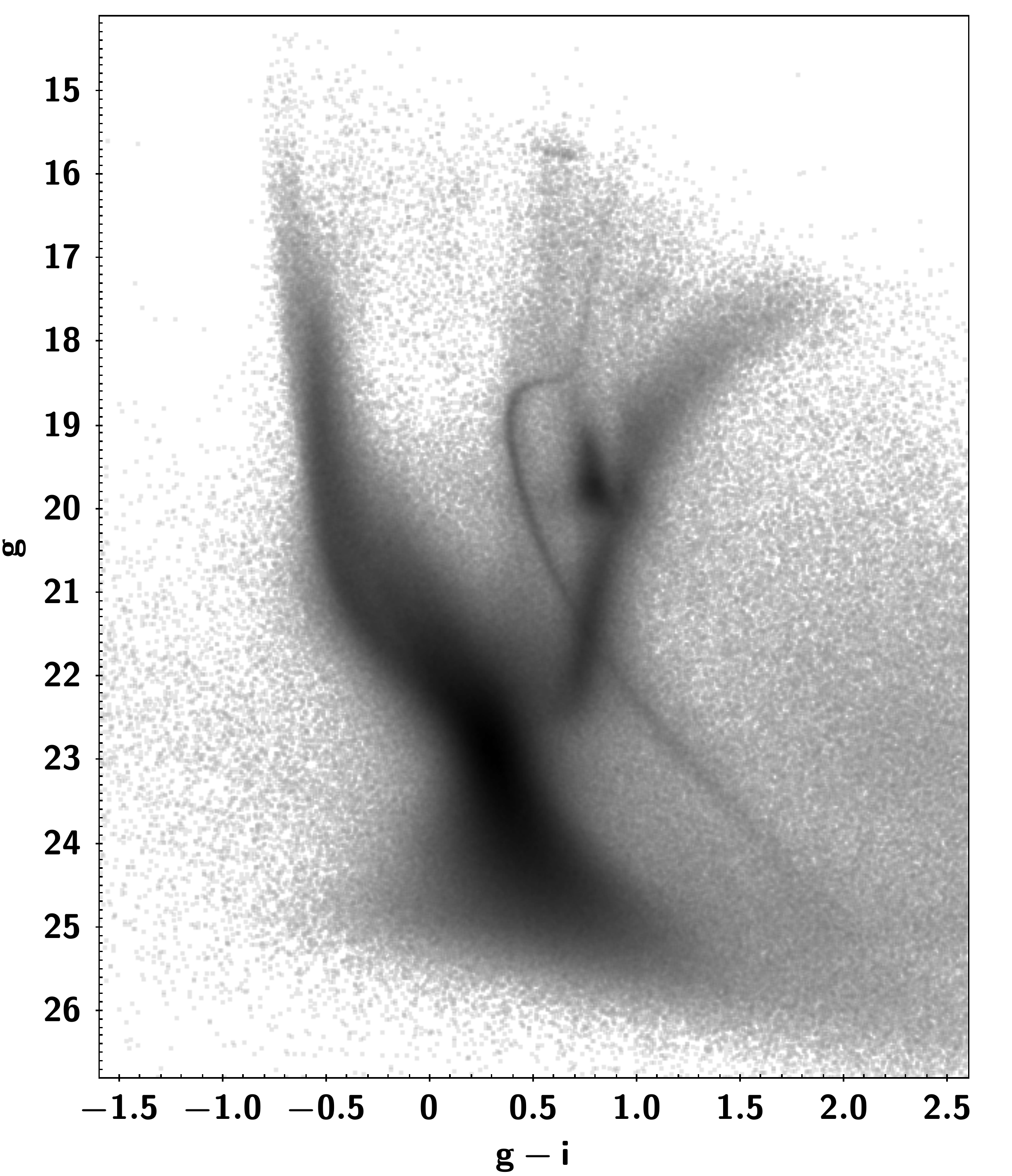}\includegraphics[width=0.5\textwidth]{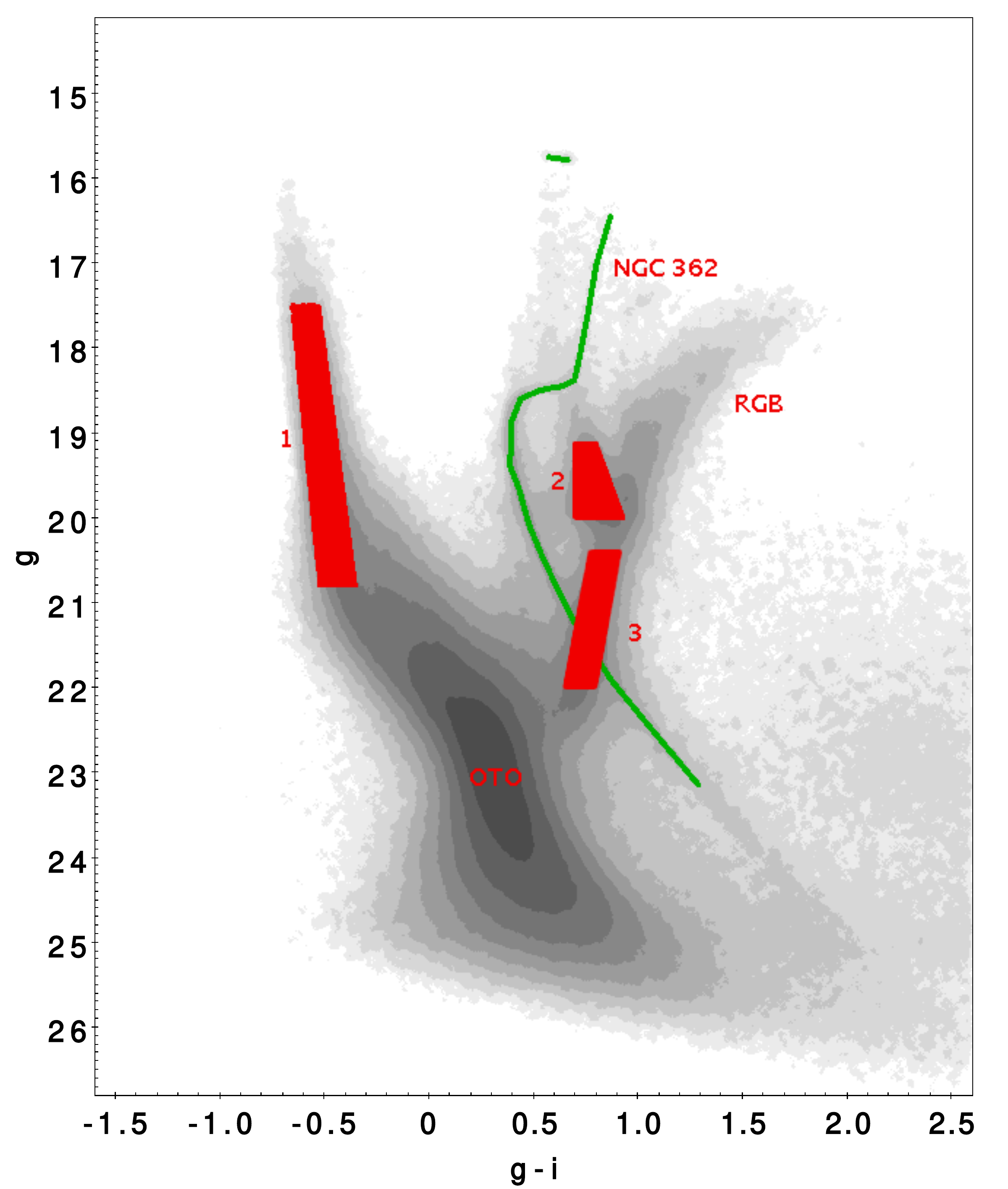}
        
    \caption{\emph{Left panel:} Colour-magnitude diagram of $\sim 2.5$ million
stars in SMASH Fields 9, 14, and 15 in the outskirts of the SMC.
Besides the strong features due to the SMC populations, the narrow main
sequence, sub-giant branch, and red giant branch of the globular
cluster NGC 362 are also
visible in this diagram. \emph{Right panel:} Density countours of the right panel highlighting the densest parts of the CMD, including the old turn-off (OTO) and the RGB.  The red boxes isolate other features of the SMC stellar population: (1) Upper main sequence, (2) red clump, and (3) lower part of the red giant branch.}
    \label{fig-cmd}
\end{figure*}

SMASH (Nidever et al.\ 2017) was carried out using the Dark Energy
Camera (DECam, Flaugher et al. 2015) on the 4-m Blanco Telescope at
Cerro Tololo Inter-American Observatory, Chile. The survey, covering
480 deg$^2$ of the sky, aims to explore the extended stellar
populations of the Clouds and their interaction history. For our current work we use a small subset
of the SMASH observations.  The region of the observed shell in the
Canon images is well covered by SMASH in Fields 9 (central coordinates
(J2000): $\alpha=$01:01:27.40, $\delta=$-70:43:05.51), 14
($\alpha=$01:20:26.78, $\delta=$-71:15:32.76), and 15
($\alpha=$01:24:33.52, $\delta=$-72:49:30.00), which are part of a
ring of fields surrounding the central part of the SMC.  Each DECam field 
covers a FoV of 3 deg$^2$. Each of these fields was
observed in the {\it ugriz} bands, with deep observations of 999s
in $u$, $i$,
and $z$, and 801s in $g$ and $r$. In addition, for each field we
obtained three shorter exposures (60s) with large offsets to tie all
the chips together photometrically and to allow
us to cover some of the gaps between the CCDs.  Details on the data processing
and photometry can be found in Nidever et al. (2017). 

The photometry of the DECam images was carried out using the photometry program DAOPHOT
(Stetson 1987) as incorporated into IRAF.  In order to identify point
sources in the photometry we imposed the following cuts in the DAOPHOT
parameters: $-1.0 < {\rm SHARP} <1.0$ and $\chi^2<3.0$. We also
required that the Sextractor (Bertin \& Arnouts 1996) stellar
probability index was ${\rm PROB} >0.8$.  Since the Fields 9, 14, and
15 cover a contiguous part of the sky and overlap with each other, we
took care to avoid double entries in the final combined catalog.  This leaves
us with 2,651,378 stars.

\section{A shell in the outer region of the SMC}

The images of the SMC obtained with two different Canon 200 lenses
(Fig.1B and Fig. 2) revealed a clear shell-like feature situated at $1.9\degr$ from the
centre of the galaxy, just above
the bar of the SMC (Figure 2). The shape of this over-density is similar to those
of the tidal features recently reported in the outskirts of the northern LMC
(Mackey et al. 2016; Besla et al.  2016) and the SMC (Pieres et al.
2017)\footnote{Unfortunately, the position of this shell in the SMC
halo (situated at 8 $\deg$ north from the SMC) is outside of the sky area covered
in our photographic survey,
including the Canon 50 images described in Besla et al. (2016).}. In
this section, we explore the spatial extent, structure, and stellar
populations by analysing the deep photometry obtained in the SMASH
survey.

\subsection{Spatial distribution as traced with different stellar populations}

\begin{figure*}
	\includegraphics[width=0.49\textwidth]{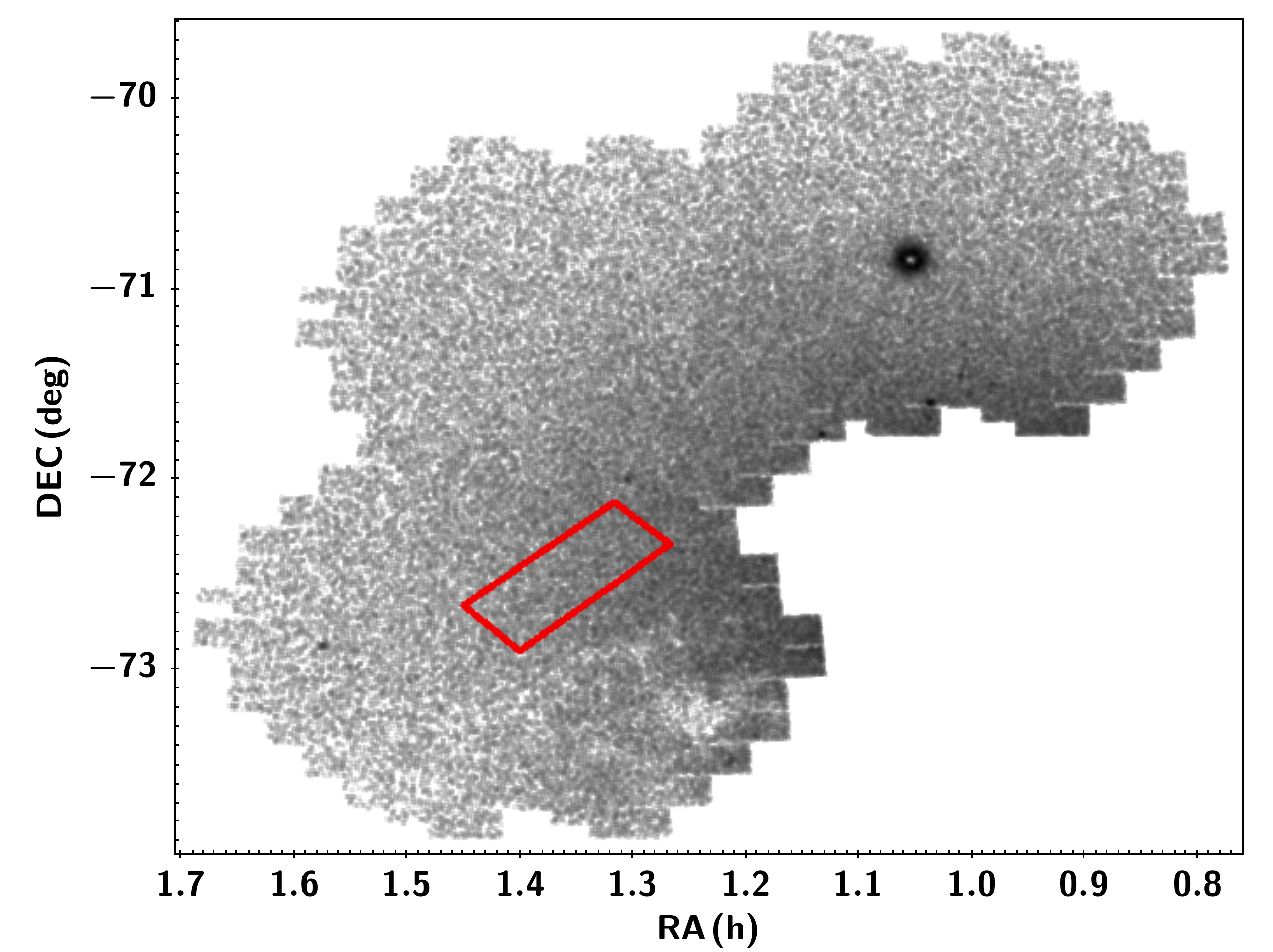}
	\includegraphics[width=0.49\textwidth]{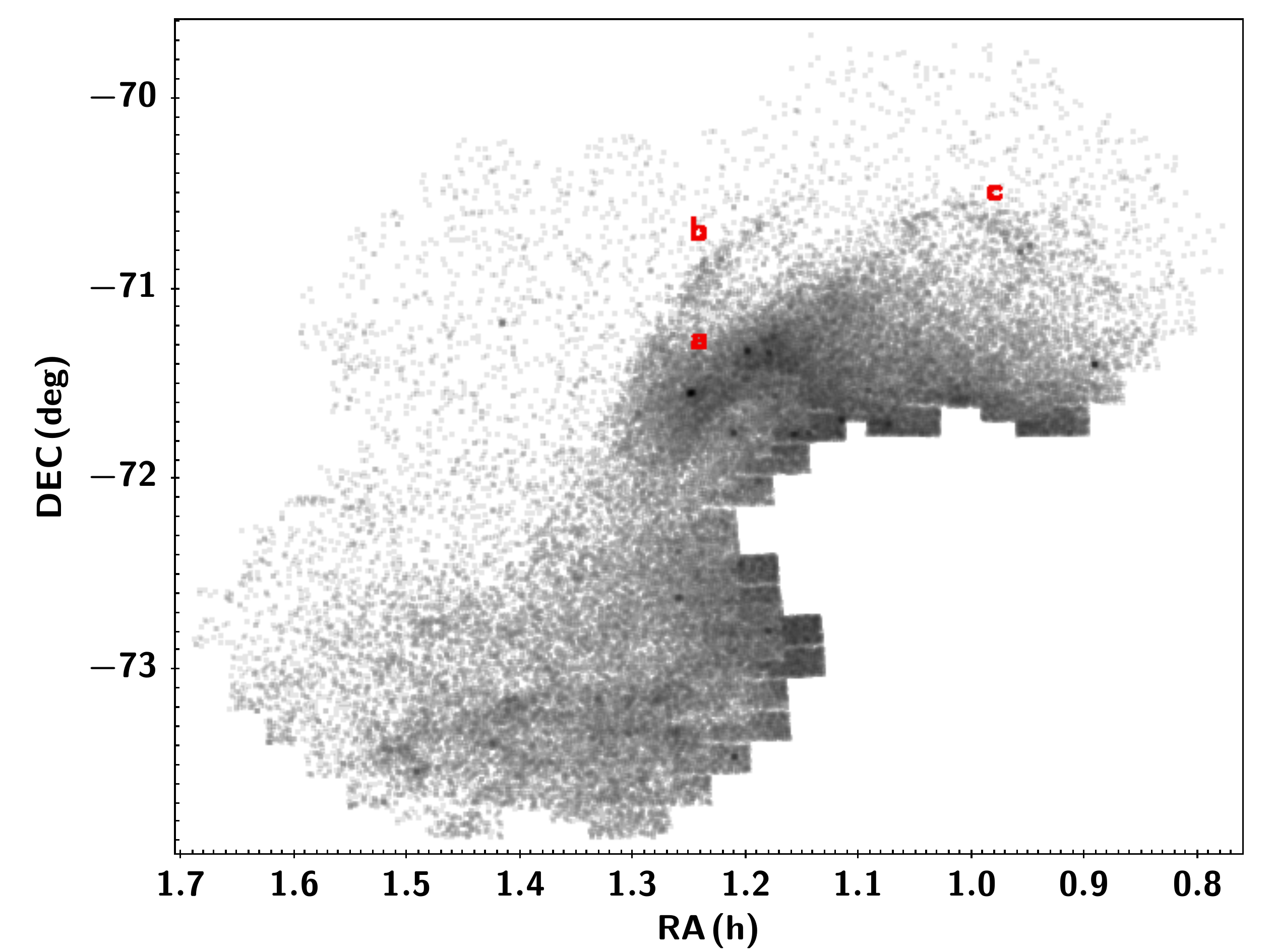}
    \caption{\emph{Left panel:} Density map of stars in boxes 2 and 3 of Figure~\ref{fig-cmd} corresponding to stars on the RC and RGB. The position of the control field ConF1 (see Sec. 3.2) is marked with a red quadrilateral. \emph{Right panel:} Density map of stars in box 1 of Figure~\ref{fig-cmd} corresponding to
stars on the upper MS. The shell-like featured detected in the Canon 200 images is marked with label {\it a}. The map reveals two new features: a spiral arm-like structure (label {\it b}) and an outer, small over-density of young stars (label {\it c}) in the vicinity of the globular cluster NGC 362. }
    \label{fig-maps}
\end{figure*}

Figure~\ref{fig-cmd} shows a $g$ versus $(g-i)$ color-magnitude diagram (CMD) of our final catalog. The features of the SMC are clearly seen in this rich diagram, and it shows the wide range of ages of the stellar
populations that are typical for this kind of dwarf irregular galaxy.
The main sequence (MS) is very extended, tracing the SMC's young
populations.  The sub-giant branch (SGB) and a wide red giant branch
(RGB) are produced by older populations with typical ages of 1 Gyr or
more. The vertical red clump and the red clump (RC) at
$19.2\lesssim g \lesssim 20$ are additional obvious features in the
diagram, which trace intermediate-age populations.  For a detailed
exploration of the star formation history of the SMC from wide-field,
multi-colour, ground-based imaging we refer the reader to Rubele et al. (2018)
and from very deep HST imaging to Cignoni et al.\ (2012, 2013).

In a first approximation, we selected stars in the main features described above, which are labeled in Figure~\ref{fig-cmd} as boxes 1 (upper MS), 2 (RC) and 3 (lower part of the RGB). Since all of these features are located in the brighter part of the CMD, the spatial coverage is uniform. This is not necessarily true in the lower part of the CMD since the deepest SMASH exposures were not dithered and hence the gaps among CCDs would be noticeable. 

To investigate the nature of the potential shell, we constructed
density maps of the region based on stars in different parts of the
CMD, as shown in Figure~\ref{fig-maps}. In the left panel we
show a map containing stars in the RC and RGB of the SMC ($\sim 100,000$ stars from boxes 2 and 3 of Figure~\ref{fig-cmd}). The distribution 
of these stars, which trace the intermediate-age and old populations of
the SMC, is quite uniform in the region, showing only a smoothly
increasing density toward the centre of the SMC located in the lower right part of the panel, at $\alpha_{J2000}=00$:52:38, $\delta_{J2000}=-72$:48:01. The globular cluster NGC 362 (top right position) is clearly visible in this map since part of its MS population overlaps with our selection box for the SMC RGB stars.  

The smooth distribution of the density of stars in this map suggests
that there is no significant extinction in this part of the sky that may
cause an apparent shell-like feature. This is also supported by the extinction map
of the SMC of Haschke et al.\ (2011, their fig.\ 4), which shows
generally low extinction in the SMC and values of $E(B-V) \simeq
0.05$ mag in the northeastern part of the SMC.  The lack of an over-density
in the upper panel of Figure~\ref{fig-maps} is furthermore consistent
with Zaritsky et al.\ (2000) who also found no evidence of the
northeastern over-density in their stellar density maps obtained using
giants and red clump stars as tracers (see their Fig.\ 3).

\begin{figure*}
\includegraphics[width=0.8\textwidth]{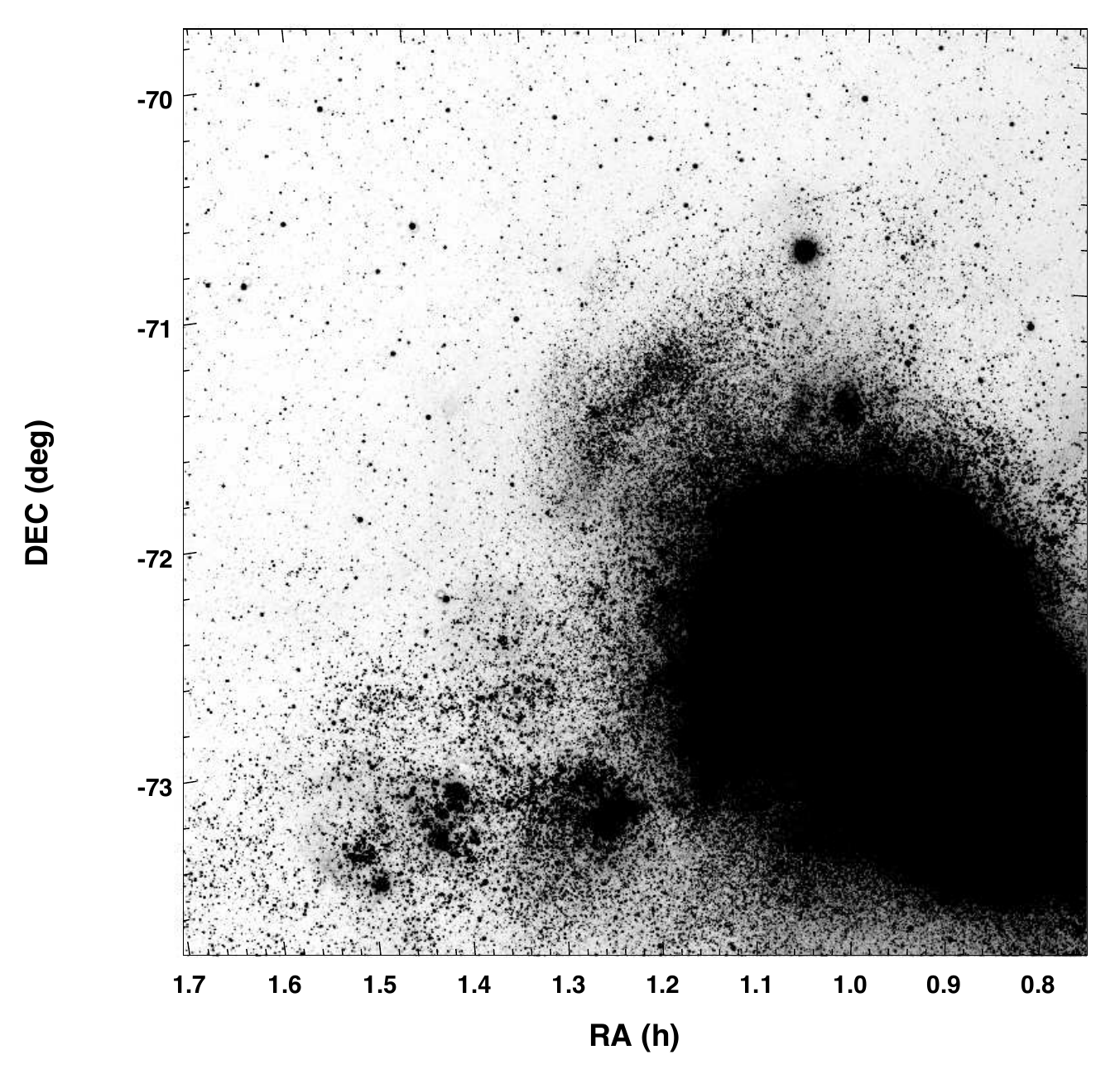}
\caption{An ultraviolet image of the SMC covering a similar field to that shown in Fig.~\ref{fig-maps}. The shell features seen by the contribution of young stars are obvious in this near-ultraviolet image of the SMC taken by the Galaxy Evolution Explorer (GALEX, $\lambda_{eff} \sim 2271 \AA, 17871 -- 2831 \AA$, FWHM $\sim$5.5"). The field has been smoothed with a Gaussian kernel of FWHM 12 pixels (18"), and scaled to highlight the low-surface-brightness shell features. The region shown consists of 65 fields with a median exposure time of 866 seconds and a signal-to-noise of $\sim$5 at 23 AB magnitude (Seibert \& Schiminovich, in preparation).}
\label{fig-galex}
\end{figure*}

The region has a very different appearance when plotting only young stars belonging to the upper MS of the SMC population, as selected by Box 1 in Figure~\ref{fig-cmd}. In this map, made with $\sim 70,000$
stars, the shell is clearly visible (labeled {\it a} in the right panel of Fig~\ref{fig-maps}) and a rich structure is associated
with it. Besides the main shell, we find a spiral-arm-like feature of young stars attached to the shell (labeled {\it b} in Fig~\ref{fig-maps}) and a separated small arc situated $\sim$ 30$\arcmin$ West from the globular cluster NGC 362 (labeled {\it c} in the right panel of Fig~\ref{fig-maps}). This last feature has no counterpart in the old population maps, so we conclude that it is not part of a tidal tail from this cluster (see also Carballo-Bello 2019). Instead, it is very interesting that two of the young open clusters discussed in Sec. 3.3 (see Table \ref{tab-cluster}) are embedded in the Southern extreme of this small structure (see Fig.\ref{fig-cluster}). The structures  {\it a} and {\it b} (Figure 4, right panel) are also clearly visible in the GALEX image
plotted in Fig.~\ref{fig-galex}, which traces mainly hot young stars, showing an excellent agreement with the position and morphology of these
features as traced in our stellar density maps. 

\begin{figure*}
	\includegraphics[width=0.6\textwidth]{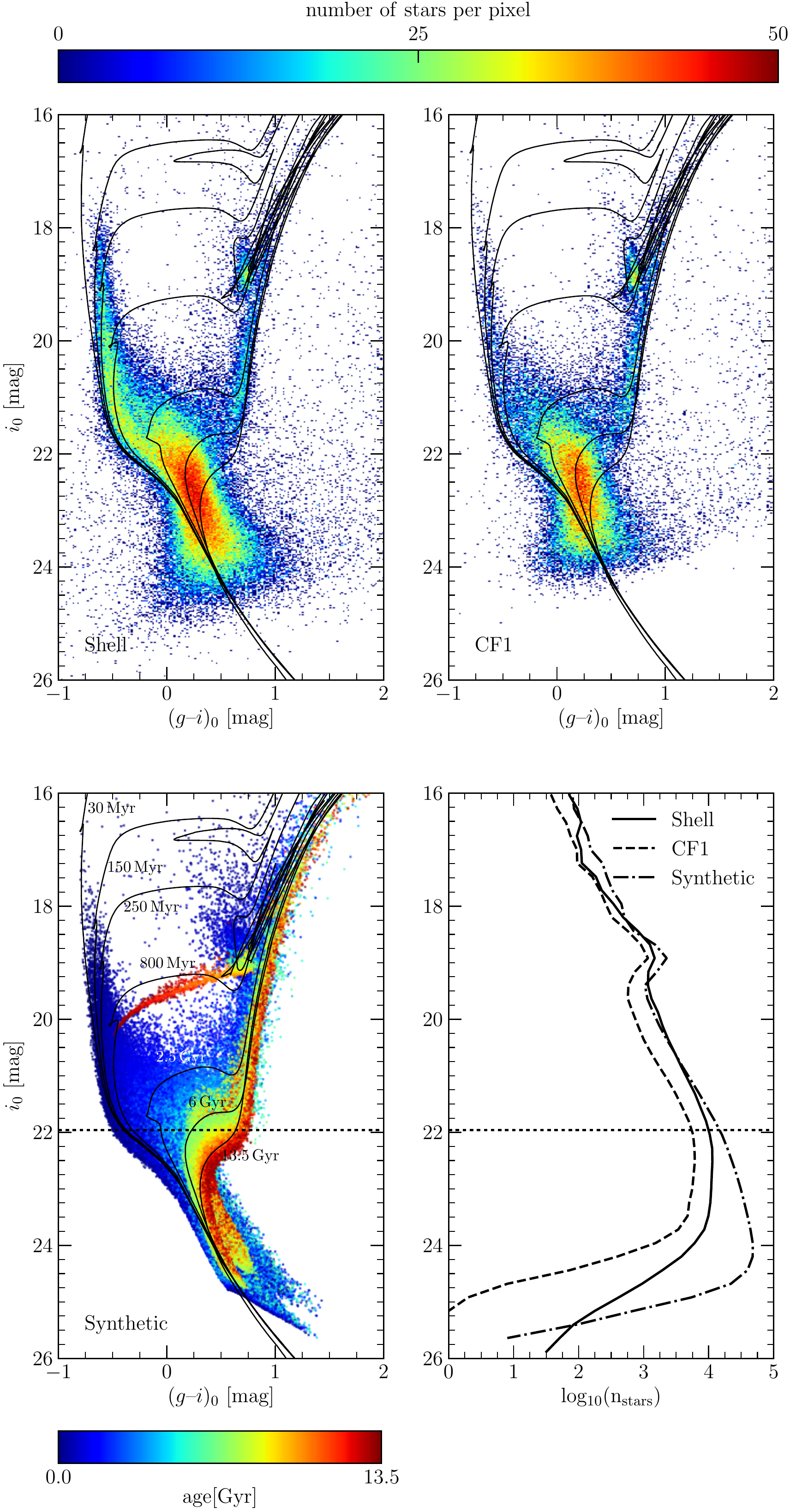}\\

    \caption{\emph{Top row:} De-reddened $i_{0}$ vs. $g-i_{0}$ Hess diagrams of the shell (left) and control field 1 (ConF1, right). The solid lines are BaSTI isochrones with ages of 30, 150, 800 Myr, 2.5, 6, and 13.5 Gyr shifted assuming a distance modulus of 18.96 mag. The metallicity of the oldest isochrone is Z=0.001 dex, whereas we have adopted a metallicity of Z=0.002 dex for the younger isochrones (see text). \emph{Bottom left:} $i_{0}$ vs. $g-i_{0}$ synthetic CMD computed using the BaSTI stellar population synthetic program web tool for a constant SFH from 13.5 Gyr to 30 Myr ago. The solid lines are the same isochrones as shown in the top row. The horizontal dashed line corresponds to our absolute magnitude cut of $M_{i}=3$ mag for deriving the colour function (see text). \emph{Bottom right:} $i_{0}$ luminosity functions of the shell, ConF1, and the synthetic CMD. The horizontal dashed line again corresponds to $M_{i}=3$ mag and demonstrates that at fainter magnitudes completeness starts to become an issue for the shell and ConF1 CMDs.}
    \label{fig-sfh}
\end{figure*}

Our CMD selection also includes a known streamer of young stars into the Bridge at $\delta \sim$ -73.3 $\deg$, which is also evident in the East side of
the SMC in our Canon 200 image in Figure~\ref{fig-canon}. This is also the origin of
some negative features imprinted in this region in the left panel of Figure~4 at $\delta \sim$ -73.3 $\deg$.
This is due to the incompleteness in these more crowded areas, since old red stars are harder to detect near bright blue star formation regions.

\subsection{Stellar content}

Fig.~\ref{fig-maps} indicates that the shell contains a rich population of young MS stars with a range of ages, which is much less prominent in the surrounding areas. To study the characteristics of the stellar populations present in the shell, we will use de-reddened $g_0$, $(g-i)_0$ CMDs and the corresponding colour functions (CF). Given the spatially variable nature of the foreground Galactic reddening, we use the reddening maps of Schlegel, Finkbeiner, \& Davis (1998) and combine these with the revised extinction coefficients of Schlafly \& Finkbeiner (2011) to de-redden each star individually.  We adopt the recommended ``mean'' distance modulus of 18.96 to the SMC as advocated by de Grijs \& Bono (2015).

Given the areal coverage of the SMASH survey in conjunction with
different environmental effects in each field across the survey --
such as distinct SFHs, differential reddening,
crowding effects, depth, etc. -- we opt to compare the de-reddened $g_0$,
$(g-i)_0$ CMD and CFs of the shell region against two different ``control'' fields: ConF1 is located close to the shell (see left panel in Fig.~\ref{fig-maps}), and ConF2 is on the
opposite side of the SMC (not shown in Fig.~\ref{fig-maps}). We aim to investigate the characteristics of the stellar content in each field, and identify differences in
the stellar populations between them. 

In Fig.~\ref{fig-sfh} we plot the CMD of the stars in the shell box (upper left panel) and in ConF1 (upper right panel). Box boxes have the same area. Isochrones from the BaSTI version 5.0.1 library (Pietrinferni et al.\ 2004\footnote{\url{http:/albinone.oa-teramo.inaf.it}}) of ages of 30 Myr, 150 Myr, 250 Myr, 800 Myr, 2.5 Gyr, 6.0 Gyr (Z=0.002) and 13.5 Gyr (Z=0.001) have been plotted as reference. A value of Z=0.002 is consistent with that derived from H\,{\sc ii} regions, young stars, and Cepheids in the SMC (Russell \& Dopita 1992; Romaniello et al.\ 2009; Lemasle et al.\ 2017). Although the SMASH survey utilizes the 4-m Blanco telescope and DECam imager, who filter system is described in Abbott et al. (2018), the photometric calibration relies on the use of standard stars in the Sloan Digital Sky Survey (SDSS; see Nidever et al.\ 2017) and as such we adopt the BaSTI isochrones in the SDSS {\it ugriz} system for our comparison.

From the two upper panels of Fig.~\ref{fig-sfh}  it seems that the shell sample contains a significantly higher number of luminous MS stars than
the same region of the CMD in the ``control'' field ConF1 sample, particularly in the age range 150-250 Myr. It is interesting, however, that the CMD of ConF1 contains a larger quantity of stars younger that 150 Myr, which are very scarce in the shell sample: from the comparison with the isochrones it can be concluded that very little star formation has taken place in the shell during the last 150 Myr.
A step in the density of stars in the main sequence can also be observed, both in the shell and in the ConF1 CMD at the approximate position of the 2.5 Gyr isochrone, possibly indicating an enhanced period of star formation at intermediate ages.  This enhancement is well-known and corresponds to a common LMC/SMC burst epoch at about 1.5-3 Gyr ago (e.g. see Harris \& Zaritsky 2002; Weisz et al. 2013).

The larger number of young, bright objects in the shell results in a $g$-band surface brightness for the shell sample that is more than 0.5 mag~arcsec$^{-2}$ brighter than the ConF1 control field sample  (cf.\ $\mu_{g,shell} = 25.81\pm 0.01$ and  $\mu_{g,CF1} = 26.68 \pm 0.01$ mag~arcsec$^{-2}$). We also notice this increased surface brightness in the $i$-band but with a smaller difference (cf.\ $\mu_{i,shell} = 25.55 \pm 0.01$ and $\mu_{i,cont} = 26.12\pm 0.01$ mag~arcsec$^{-2}$). For the ``mean'' SMC distance modulus of 18.96, we determine that the absolute $g$- and $i$-band magnitudes of the shell sample are M$_{g,shell} = -10.78\pm 0.02$ and M$_{i,shell} = -11.05 \pm 0.02$ mag, respectively.

The lower left panel of the Fig.~\ref{fig-sfh}  shows the CMD of a synthetic population computed using the BasTI on-line Stellar Population Synthesis Program\footnote{\url{http://basti.oa-teramo.inaf.it/BASTI/WEB_TOOLS/synth_pop2/}}, using solar scaled overshooting models.  We have assumed a constant star formation rate (SFR) from 13.5 Gyr to 30 Myr ago (the latter is the young age limit for the BasTI library), and a simplified chemical enrichment law, approximately consistent with that obtained by Carrera et al. (2008) from Ca II triplet spectroscopy, that is, [Fe/H]=-0.99 ($\sigma$=0.3) for the second half of the galaxy's life (age $<$6.75 Gyr), and [Fe/H]=-1.29 ($\sigma$=0.1) for the first half (age$>$6.75 Gyr). Other parameters of the model are a binary fraction of $\beta$=0.4 with a mass ratio q$>$0.5 and a Kroupa et al. (1993) IMF. Different colours for the synthetic stars have been used to highlight the position in the CMD of stars with different ages. The same isochrones as in the observed CMDs were plotted as reference. Finally, the lower right panel of Fig.~\ref{fig-sfh} shows the luminosity function of the shell, control fields, and that of the synthetic CMDs. These luminosity functions allow us to conclude that the observed CMDs are basically complete down to $M_i$=3 or $i_0$=21.96. The horizontal lines indicate this magnitude limit, which will be used to compute the CFs.

\begin{figure}
	\includegraphics[width=0.49\textwidth]{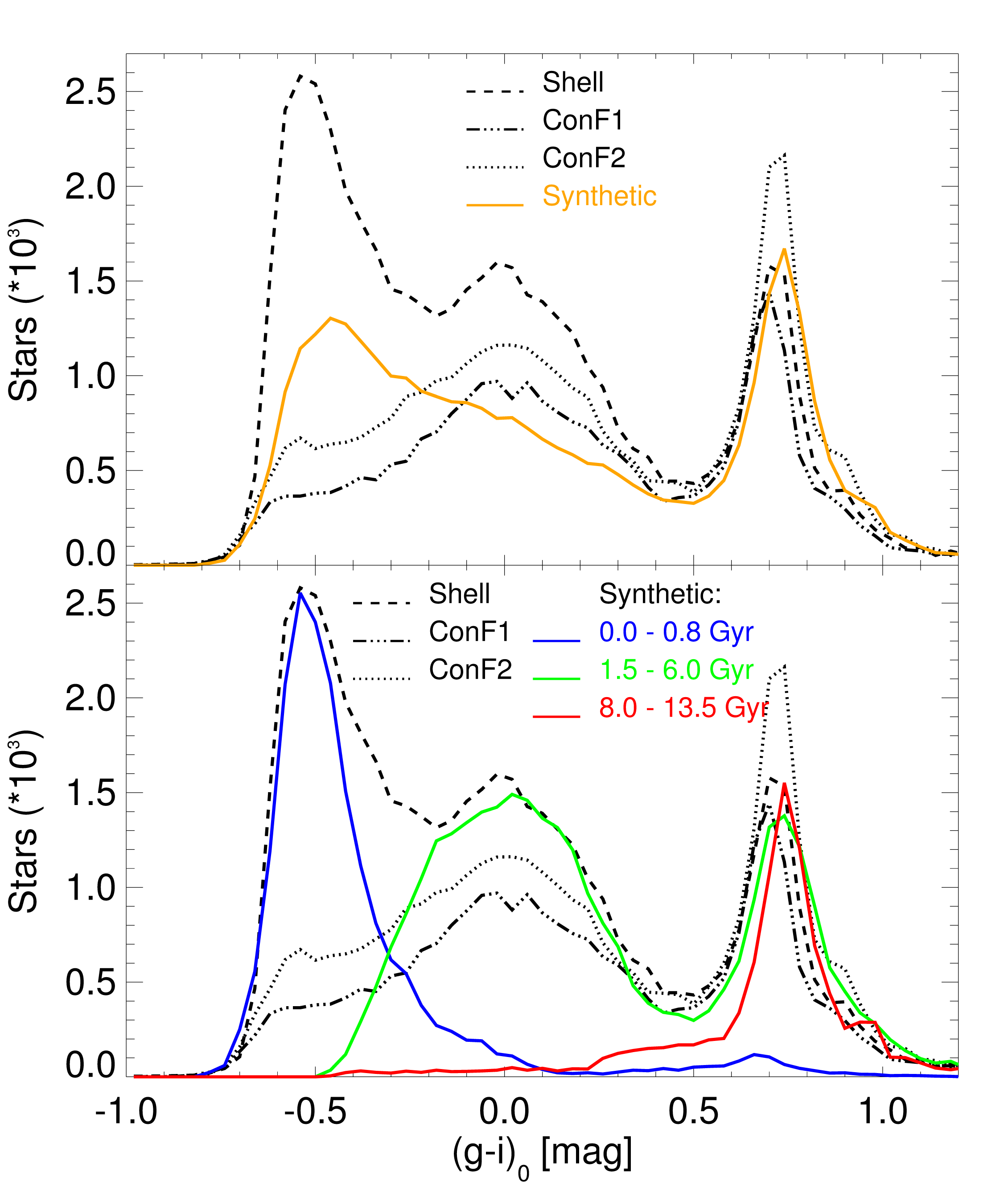}
    \caption{ \emph{Upper panel}: CFs of the shell and control fields (for $M_i<3$), with CFs for a synthetic population with constant SFH superimposed. \emph{Lower panel}: the same CFs for the observed fields, together  with the CF for three populations in different age ranges, as labelled (see text for details).}
    \label{fig-cfs}
\end{figure}

In Fig.~\ref{fig-cfs}, we use the CFs to further analyze the stellar population content of the shell and the differences to the SMC field populations at similar galactocentric radius (for an introduction to the use of the CF for stellar population analysis, see Gallart et al. 2005; see also No\"el et al. 2007 for an application to study SMC field stellar populations). We compare the shell CF with the CF of the same control field shown in Fig.~\ref{fig-maps} (ConF1) and control field ConF2 located at the opposite side of the SMC. The upper panel of Fig.~\ref{fig-cfs} shows, in black and different line types, the $(g-i)_0$ observed CFs for the shell and the two control fields mentioned above. An absolute magnitude cut of $M_i$=3 has been adopted to ensure a high level of completeness in the photometric data used to calculate the CFs (see Fig.~\ref{fig-sfh}). The orange solid line shows the CF of the synthetic CMD with a constant star formation rate at all times. The CF of the synthetic CMD was scaled such that the number of stars in a box on its red clump matches that in a box in the same location in the shell CMD.   

The lower panel of Figure~\ref{fig-cfs} displays the same observed CFs, together with CFs of the synthetic population in three limited age ranges: 0.03 to 0.8 Gyr, 1.5 to 6 Gyr, and older than 8.0 Gyr. In this case, the normalization of the synthetic CFs is arbitrary, and chosen to approximately match the features present in the observed CF of the shell. The position and width of the blue maximum in the shell CF (approximately centered at $(g-i)_0 \simeq -0.5$) suggests a very recent star formation event. The second maximum approximately centered at $(g-i)_0 \simeq 0$ is well reproduced by a population spanning the age range 1.5 to 6 Gyr. The third maximum corresponds to the position of the red clump. These results on the age range that is contributing to each feature of the CF are in good agreement with the conclusions we reached from the comparison of the CMDs with isochrones.  The comparison of the observed CFs indicates significant differences in the stellar content of the shell and the two control fields. The very prominent blue maximum in the shell CF has very low counts in both ConF1 and ConF2. The difference in the height of the second, intermediate colour maximum in the shell and the control fields is not striking, but a difference is nonetheless evident.  Taking into account the information on the lower panel of Fig.~\ref{fig-cfs} regarding the ages contributing to each feature on the CF (complemented by the isochrone information), we conclude that star formation has been very active in the shell in the last $\simeq$ 1 Gyr  and much less active in the control fields.  In the intermediate-age range 1.5-6 Gyr ago, star formation was also more active in the shell field than in the control fields. The comparison with the CF of a population with a constant star formation rate (orange line in the upper panel of Fig.~\ref{fig-cfs}) discloses that star formation in the shell field has been enhanced (compared to an average constant SFH) in the second half of the galaxy's life, while star formation in both control fields has been depressed in the last $\simeq$ 1 Gyr compared to a constant star formation rate. 
  
Our analysis of the total stellar content toward the shell field, through the comparison of the CMD with isochrones and the comparison between observed and synthetic CFs, reveals a very complex stellar population, with stars of all ages contributing to the CMD of the shell field. A strong enhancement of the recent star formation rate is, however, indicated by the height of the bluest peak of the CF. Even though a broad age range seems to be necessary to reproduce the width of that peak, the details of this age range and the precise age composition cannot be accurately constrained by this kind of simple analysis. A full SFH derivation, which is beyond the scope of this paper, would be necessary. Additional considerations brought up by the analysis of the young cluster (see Sec.3.3) and Cepheid population (see Sec. 3.4) in that area of the SMC indicate that a conspicuous burst of star formation around 150-250 Myr contributed outstandingly to the young population of the shell and may be the origin of the features that identify the shell morphology and that are visible in the right panel of Figure \ref{fig-maps}.
  
\subsection{Young Star Clusters}

\begin{table}
\begin{center}
\begin{tabular}{l|c|c|c|r}
\hline
Cluster Name & RA & DEC & log(age/yr) & Ref. \\
\hline
  B88 & 14.233 & -70.773 & 8.10 & 1\\
  HW33 & 14.346 & -70.809 & 8.10 & 2\\
  B139 & 17.617 & -71.561 & 8.30 & 1\\
  HW64 & 17.687 & -71.338 & 8.25 & 1\\
  IC1655 & 17.971 & -71.331 & 8.30 & 1\\
  IC1660 & 18.158 & -71.761 & 8.20 & 1\\
  L95 & 18.687 & -71.347 & 8.30 & 1\\
  NGC458 & 18.717 & -71.550 & 8.15 & 3\\
  HW73 & 19.108 & -71.326 & 8.15 & 1\\
\hline
\end{tabular}
\caption{Young Star Clusters in the SMC shell. RA and DEC are in J2000 system. References: (1) Glatt et al. 2010, (2) Piatti et al. 2014, (3) Alcaino et al. 2003.}
\label{tab-cluster}
\end{center}
\end{table}

\begin{figure}
	\includegraphics[width=0.49\textwidth]{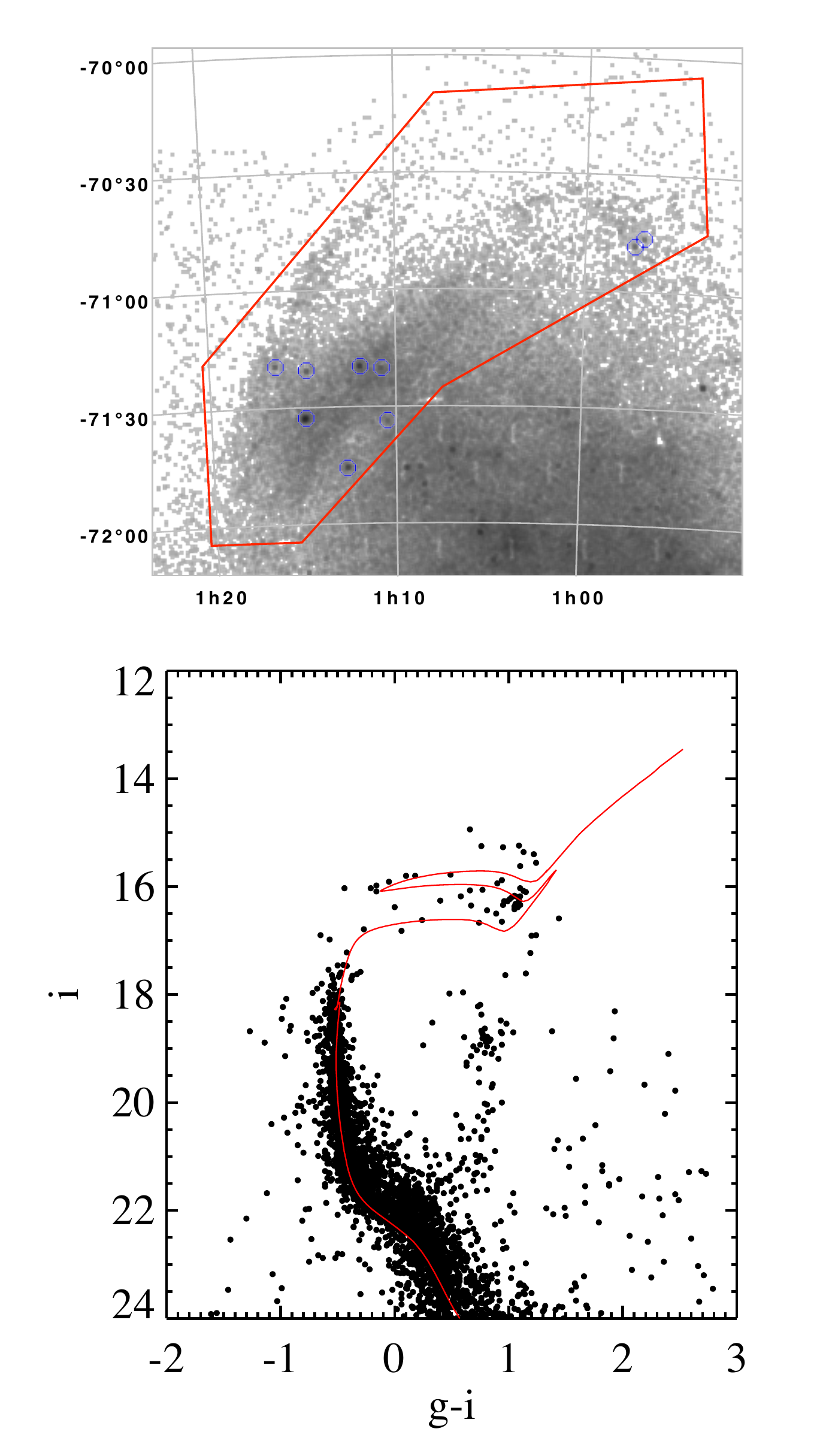}
\caption{\emph{Top panel}: MS stellar density map showing the locations of nine young ($<$1 Gyr) star clusters (blue circles) associated with the shell feature, defined here spatially by the red polygon. \emph{Bottom panel}: The combined CMD of the nine young star clusters, showing their collective compatibility with a single age of $\sim$160 Myr.  We overplot a $\sim$160 Myr PARSEC isochrone (log(age/yr)=8.2, Z=0.002, $A_V$=0.25 mag, (m-M)=18.96) for reference.}
    \label{fig-cluster}
\end{figure}

A number of young star clusters lie spatially coincident with the northeastern shell feature, with the most prominent clusters appearing as visible over-densities in the upper MS star count map in Fig.~\ref{fig-maps}. We identify nine young ($<$1 Gyr) star clusters from the Bica et al. (2008) catalog that lie within a $\sim2.2$ square degree search region. We display their location in Figure \ref{fig-cluster} (upper panel), and list their properties in Table 1. For this census, we exclude six older (age$>$1 Gyr) clusters from further discussion, and ignore 5-10 diffuse associations and low significance catalog entries. This sample features three relatively massive clusters ($\sim10^4 M_{\sun}$; NGC458, IC1655, IC1660), as well as six other less massive systems\footnote{ The association of three clusters with the position of this feature was previously mentioned by Brueck \& Marsoglou (1978). However, only NGC 458 and L90 actually fall on the shell, while HW62 is clearly offset off the stellar over-density towards the SMC center. The ages assigned to the clusters by these authors, based on photographic plates, have been revisited with our modern observations yielding 160 Myr (instead of 20-50 Myr).}.

A striking feature of this young cluster sample is their uniformity in age.  As determined from isochrone comparisons to observed CMDs (Alcaino et al. 2003; Glatt, Grebel \& Koch 2010; Piatti 2014), the cluster ages pile up around $\sim$ 160 Myr [log(age/yr)=8.2$\pm$0.1] and appear consistent with this single age, within current fitting uncertainties. We demonstrate the agreement with a single age by over-plotting an isochrone with log(age/yr)=8.2 on top of a summed SMASH CMD created for the nine cluster sample (see lower panel in Fig.~\ref{fig-cluster}) . This synchronization in cluster ages suggests (or, further reinforces the conclusion) that an important enhancement of the star formation rate took place $\sim$ 160 Myr ago. This epoch broadly agrees with the SMC-wide peak of cluster ages observed by Glatt et al. (2010) as well as the putative age of the most recent LMC-SMC interaction.

\subsection{Cepheids}

\par
\begin{figure}
\includegraphics[width=0.49\textwidth]{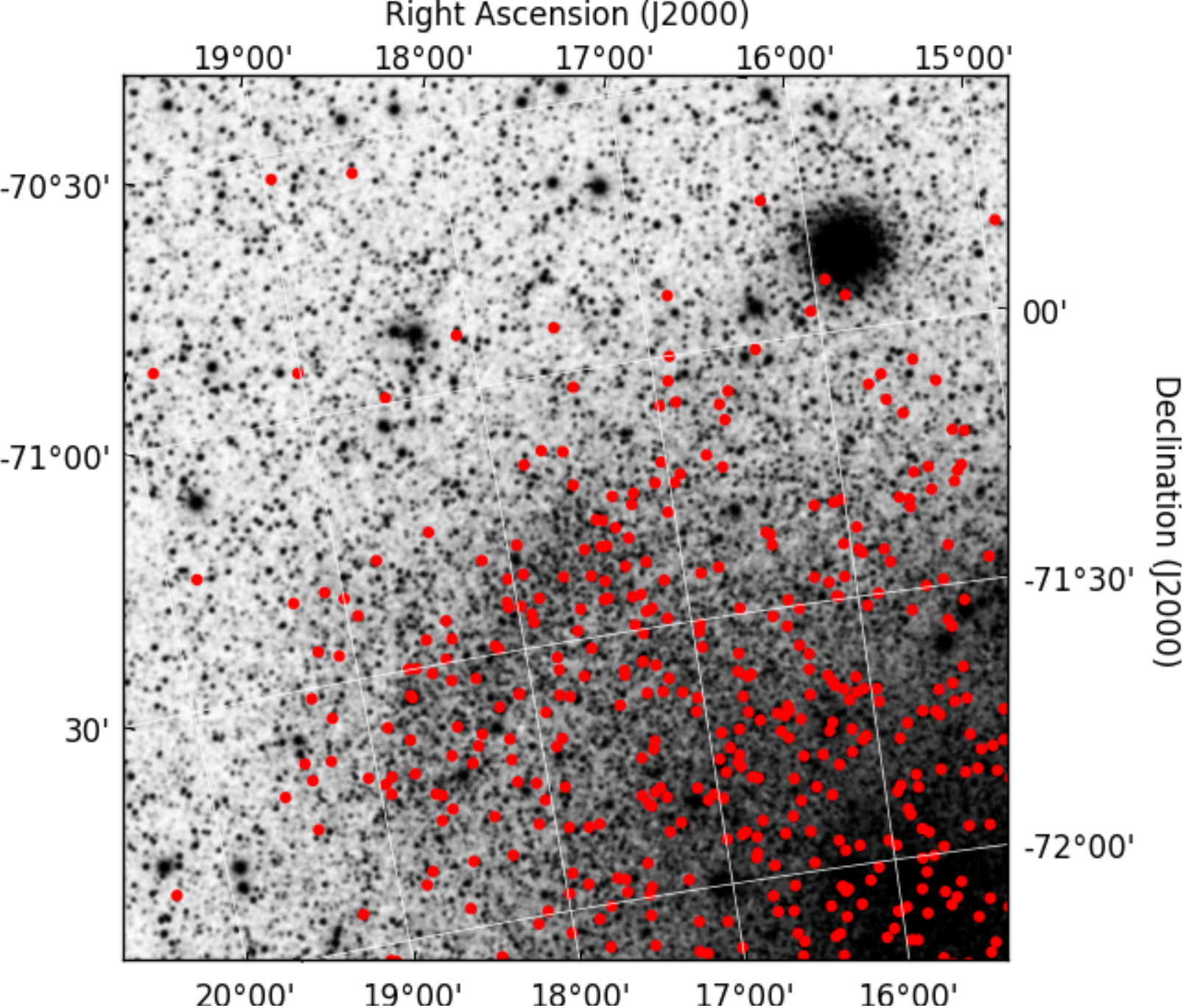}
\label{ceph}
\caption{Classical Cepheids in the OGLE database (red dots) over-plotted on the shell region.}
\label{fig-ceph1}
\end{figure}

 OGLE IV  (Udalsky et al. 2015; Soszy\'nski et al. 2016) uncovered almost 5,000 classical Cepheids in the SMC and several tens of them spatially overlap with the shell feature (see Fig.~\ref{fig-ceph1}). Since Classical Cepheids are young supergiants, their presence in quite large numbers underlines significant star formation in the last few hundred Myr. 
\par
\begin{figure}
\includegraphics[width=0.45\textwidth]{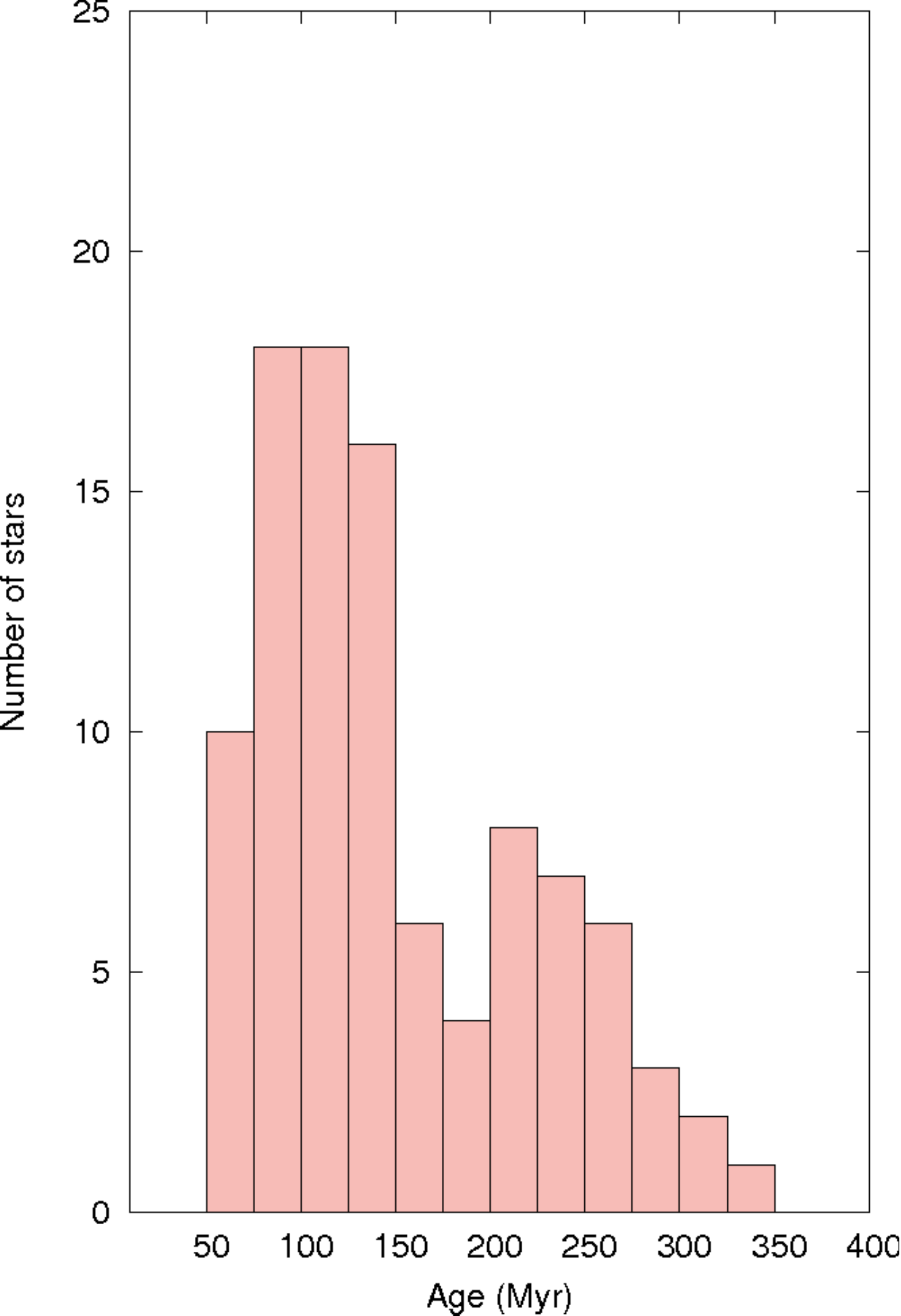}
\label{ceph_ages}
\caption{Age distribution for Cepheids located in the vicinity of the shell. Ages have been computed with the period-age relations for fundamental and first overtone pulsators of Bono et al. (2005) at Z=0.004.}
\label{fig-ceph2}
\end{figure}

 Cepheid ages can be computed for individual stars using period-age relations derived from population models (for instance, Bono et al. 2005). Using these relations, the ages vary from $\approx$15 to $\approx$500 Myr\footnote{Models that include rotation during the Main Sequence phase lead to Cepheid ages increased by 50 to 100\%, depending on the period (Anderson et al. 2016)} for SMC Cepheids. The age distribution of SMC Cepheids is known to be bimodal, with two peaks at $\approx$110--130 Myr and $\approx$220--230 Myr separated by a minimum at $\approx$150 Myr (e.g., Inno et al. 2015; Subramanian et al. 2015; Jacyszyn-Dobrzeniecka et al. 2016; Ripepi et al. 2017). The former peak has been associated with star formation triggered by the most recent interaction between the LMC and the SMC. Cepheids lying in the shell region span a relatively wide age range that clearly peaks at 100--130 Myr (see Fig.~\ref{fig-ceph2}). Such ages are in good agreement with current results for the MS sample and match very well the ages derived for the young clusters in the vicinity of the shell\footnote{However, it is important to take into account that the raw age distribution of Cepheids cannot be directly interpreted as an age distribution of the underlying star formation because it is convolved with both the stellar IMF and the lifetime of the star within the instability strip during which it would be identified as a Cepheid variable.}. Cepheids therefore support a scenario where the shell population is dominated by young stars formed during a recent star formation event, possibly related to the interaction between both Magellanic Clouds.

\subsection{Stellar Kinematics}
\label{sec-kin}
\begin{figure}
\includegraphics[width=0.5\textwidth]{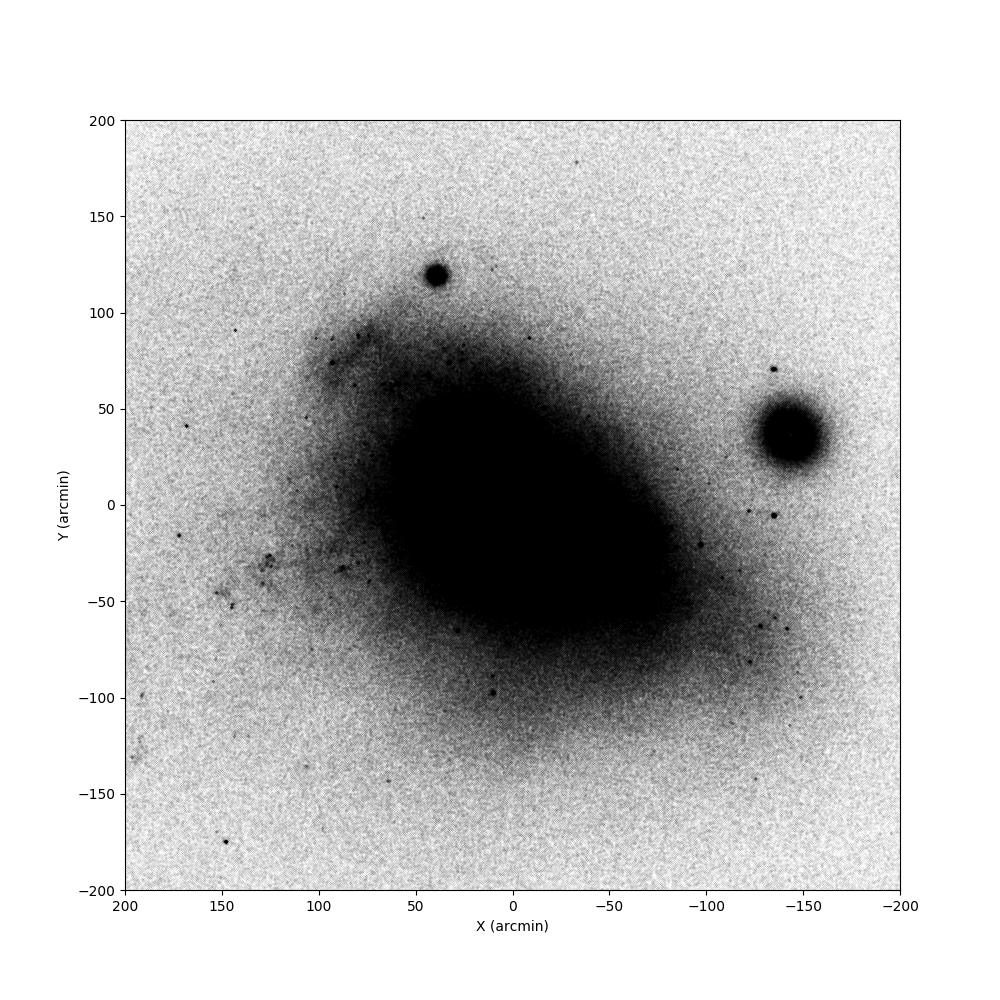}
\includegraphics[width=0.5\textwidth]{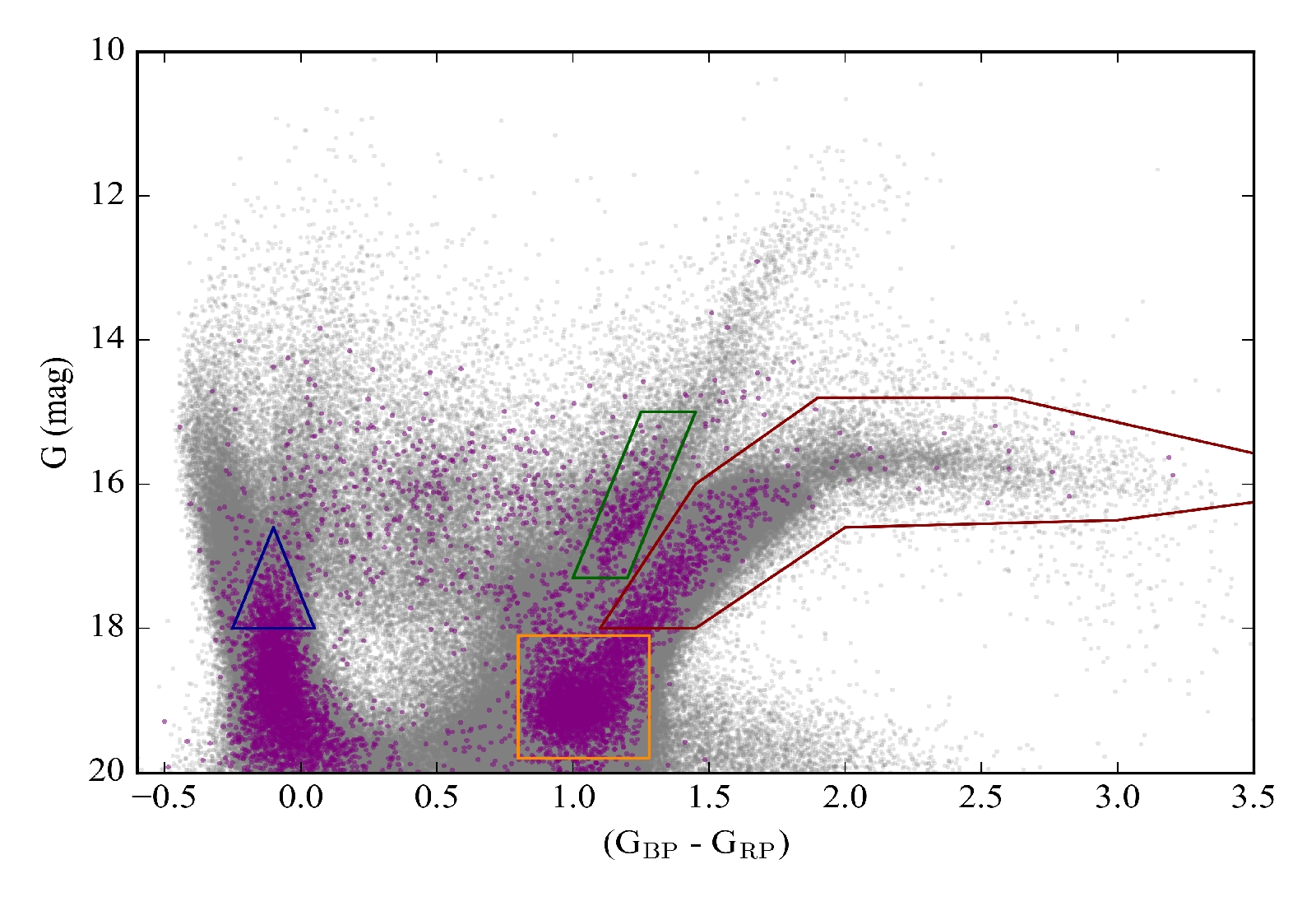}
\caption{\emph{Top panel:} Star map made with all the {\it Gaia} DR2 (Gaia Collaboration et al. 2018) sources with \textsc{visibility\_periods\_used} $\geq$ 5 and \textsc{phot\_bp\_rp\_excess\_factor} < 1.5 in a 400\,arcmin $\times$ 400\,arcmin box centered in  the SMC.
For orientation reference, the LMC is located down and to the left of the plot. The previously described shell and new features showed in Fig.4 (right panel) are clearly visible in the $Gaia$ data. \emph{Bottom panel}: CMD of an astrometrically selected sample of \textit{Gaia} sources with G $<$ 18 and $|{\mu}| < 5$ mas yr$^{-1}$ (marked in gray) with all sources selected within the shell region (the exact area selected can be seen in the top panel of Figure \ref{fig-gaia2}) over-plotted in purple.  The masks for selecting different CMD features are also over-plotted: main sequence (MS, blue), red supergiants (RSG, green), red clump (orange) and red giant branch (red).}
\label{fig-gaia1}
\end{figure}

Using the \textit{Gaia} DR2 data (Gaia Collaboration et al. 2018), proper motions for thousands of stars in the SMC outer region have recently became available. We use this catalog to investigate the kinematic signature from the shell-like feature. From the DR2 database, we select stars surrounding the SMC and apply a series of astrometric cuts. We start by applying a parallax cut of $\omega < 0.2$ mas in order to remove foreground MW stars. Next we use a cut to the renormalized unit weight error (as described in the \textit{gaia} technical note GAIA-C3-TN-LU-LL-124-01) of 1.40 and a cut for the color excess of the stars (as described in Lindegren et al. 2018 by Equation C.2). As astrometric precision has a strong relationship with the magnitude of the stars, we additionally apply a cut of $\mathrm{G} < 18$. Finally, to trim down potential MW contamination, we cut out an area with radius equal to 3 mas yr$^{-1}$ around the systemic proper motion (PM) of the SMC, which we will take to be $\mu_{\alpha^{*},0,\mathrm{SMC}}, \mu_{\delta^{*},0,\mathrm{SMC}}$ = $0.797 \pm 0.030, -1.220 \pm 0.030$ mas yr$^{-1}$ (Helmi, et al. 2018). This leaves us our final selection of stars (seen in upper Figure \ref{fig-gaia1}), where the shell-like over-density can be clearly seen (marked by the purple rectangle).

\begin{figure}
 \includegraphics[width=0.5\textwidth]{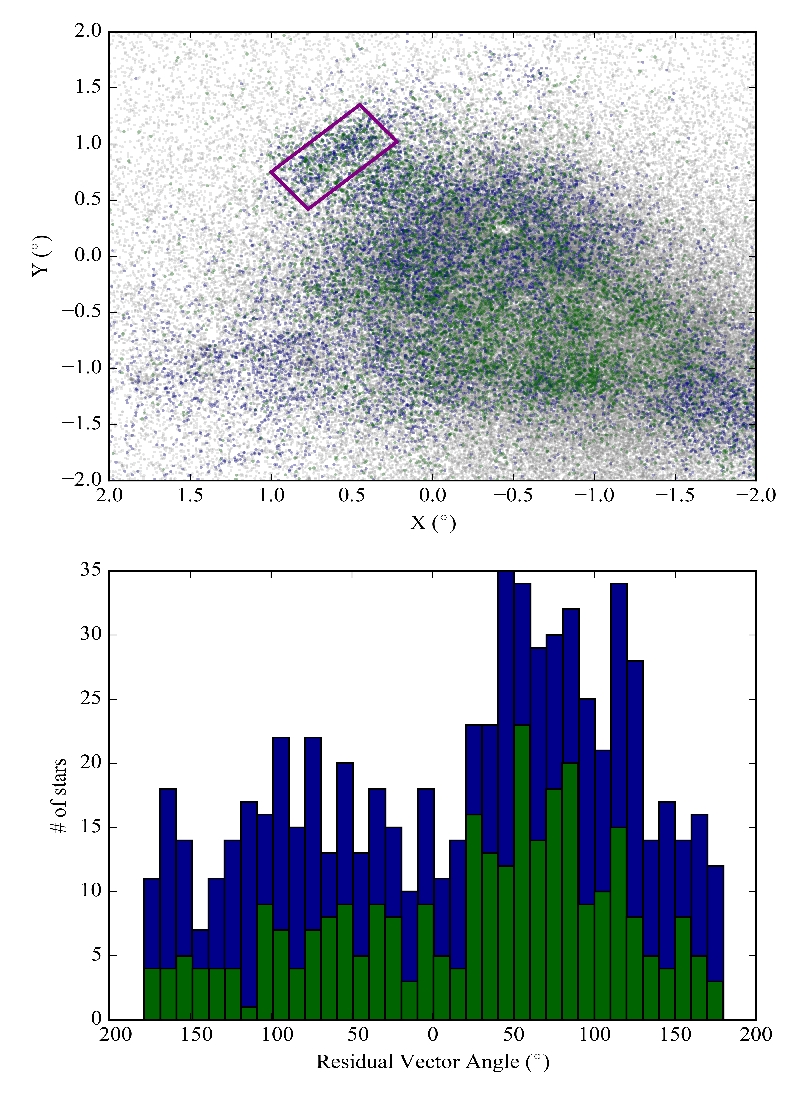}
 \caption{\emph{Top panel:} Spatial plot of our selected \textit{Gaia} sample (marked in grey), discussed in Section \ref{sec-kin}, with the main sequence (MS, blue) and red supergiant (RSG, green) sequences shown in Figure \ref{fig-gaia1} overplotted. The shell feature can be seen towards the center-left area of the plot in addition to significant substructure in the two sequences spread throughout the SMC. This feature has been marked with a purple box, and all stars that fall within the box are examined below in addition to being overplotted in the CMD for the SMC (seen in the bottom panel of Figure \ref{fig-gaia1}). \emph{Bottom panel:} Combined histogram of the residual proper motion vector angles of the MS and RSG populations after removing the systemic motion of the SMC and correcting for viewing perspective. Only stars that fall within the marked box in the top of Figure 12 are displayed, indicated by the same colors as above. The angle measurement is defined so that a residual vector pointing vertically in the spatial plot corresponds to an angle of $0^{\circ}$, and the angle increases in a counter- clockwise direction. A clear preference for a mean residual vector angle of $\sim 70-80^{\circ}$ can be seen, which roughly points radially outwards from the center of the SMC.}
   \label{fig-gaia2}
\end{figure}

The CMD of the shell region obtained from the Gaia data (bottom panel of Figure \ref{fig-gaia1}, expanded down to G $= 20$ for greater context) displays clearly the MS, the red supergiants (RSG), the RC, and the RGB features. Combining the requirement of G $< 18$ and the apparent locations of the stellar features, we created masks for each feature. The masks were then applied to the full sample of SMC stars and the resulting spatial distributions were plotted (upper Fig. \ref{fig-gaia2}). Similar to the analysis earlier in the paper, the RGB had no apparent correlation with the shell-like region, but both the MS and RSG display over-densities in the location of the shell. Correspondingly, we select these two sequences from the shell region for our kinematic analysis. 

For this portion of the analysis, we first subtract the systemic PM from the PM of each star in the shell region. As the feature is noticeably extended on the night sky, we also calculate and remove the viewing perspective for each source, as outlined in van der Marel et al. (2002). These residual proper motions are converted into a Cartesian frame assuming the kinematically-derived center of ($\alpha_{J2000}$, $\delta_{J2000}$) = (16.25$^\circ$,$-$72.42$^\circ$), using the transformations from the Gaia Collaboration (Helmi et al. 2018). The position angle (PA, which in the context of this analysis is defined as the vector angle of the residual proper motions) is calculated for each source, defined where $0^{\circ}$ points vertically upwards in the spatial plot in Figure \ref{fig-gaia2} and increases in a counter-clockwise direction. We compile all of the PAs for all sources into a histogram for easier interpretation (seen in lower panel of Figure \ref{fig-gaia2}). The PAs appear to peak around $70-80^{\circ}$, which points roughly radially outwards from the center of the SMC. The scatter from $0^{\circ}$ to $360^{\circ}$ is expected as the average residual is on the order of $\sim 0.1$ mas yr$^{-1}$ with average errors of comparable magnitude, underscoring that the large peak in PA must be a real signal. This coherent radially outward motion of stars in the SMC is consistent with prior studies of the internal kinematics of the SMC (e.g. Zivick et al. 2018).

\subsection{H\,{\sc i} and H$\alpha$  emission}

\begin{figure*}
    \includegraphics[width=1.0 \textwidth]{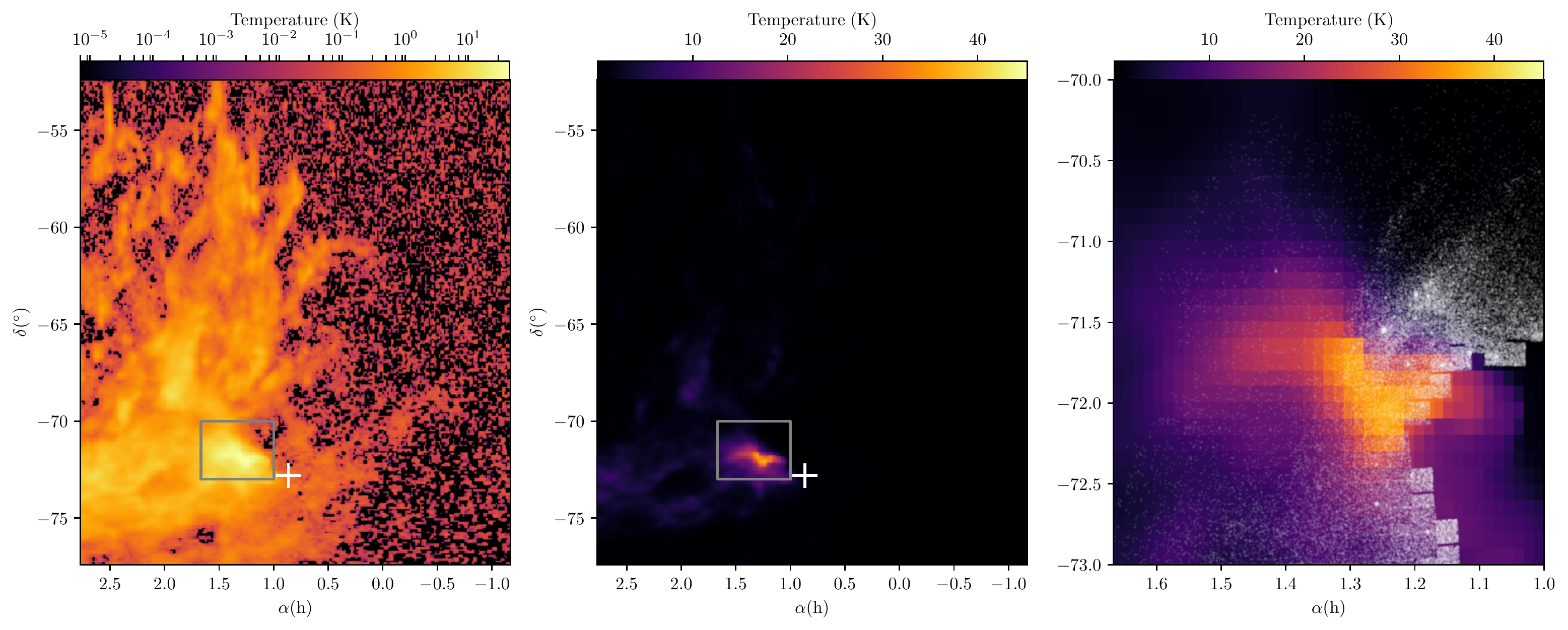}
    \caption{ \emph{(Left panel):} GASS H\,{\sc i} emission for the velocity channel $v=198$ km~s$^{-1}$, in a log scale. The center of the SMC is marked by a white plus sign. The grey rectangle in this and the following panel is zoomed in the right panel. \emph{(Central panel):} Same region as the previous panel, but showing in a linear scale the mean temperature for the mean $\pm$ 1$\sigma$ distribution in the velocity channel. A clear `Z' shape is seen with a gradient in temperature towards the center of the SMC. \emph{(Right panel):} Enlarged region (grey rectangles in the previous panels) showing the distribution of H\,{\sc i} in a linear scale over-plotted to the young main sequence stars as white dots (box 1 in Figure 3).}
    \label{V198stars}
\end{figure*}

Using the atomic hydrogen (H\,{\sc i}) emission maps from the Galactic All-Sky Survey (GASS) third release\footnote{\url{https://www.astro.uni-bonn.de/hisurvey/gass/}} (Kalberla \& Haud 2015), we scanned the velocity channels available in that survey (from $-495$ km~s$^{-1}$ out to +495 km~s$^{-1}$). The H\,{\sc i} gas emission within one square degree centered on $\alpha = 1$h10min, $\delta = -71.5{\degr})$ reaches a maximum in the channel $v = 198$ km~s$^{-1}$. more specifically, the velocities of the gas within that region are close to a Gaussian distribution, with the maximum in $v = 198.3$ km~s$^{-1}$ and a dispersion $\sigma = 13.6$ km~s$^{-1}$. This velocity is very similar to the mean velocity of the stars measured by  Evans \& Howarth (2008)(172 km s$^{-1}$) with a dispersion of 30 km s$^{-1}$. The leftmost panel of Fig.~\ref{V198stars} shows the emission for this specific velocity channel in a log scale. The second panel of Fig.~\ref{V198stars} is the distribution of the mean gas temperature bounded by a larger range of velocities (184-211 km s$^{-1}$) in a linear scale, which is very similar to the previous panel. The rightmost panel shows the gas in the respective velocity channel, in a zoomed region close to the shell discussed in this paper. The young MS stars are over-plotted on the gas density map as white dots. A visual inspection on the last panel suggests a shift between the projected location of the young stars and the gas distribution.

\begin{figure*}
\includegraphics[width=0.5 \textwidth]{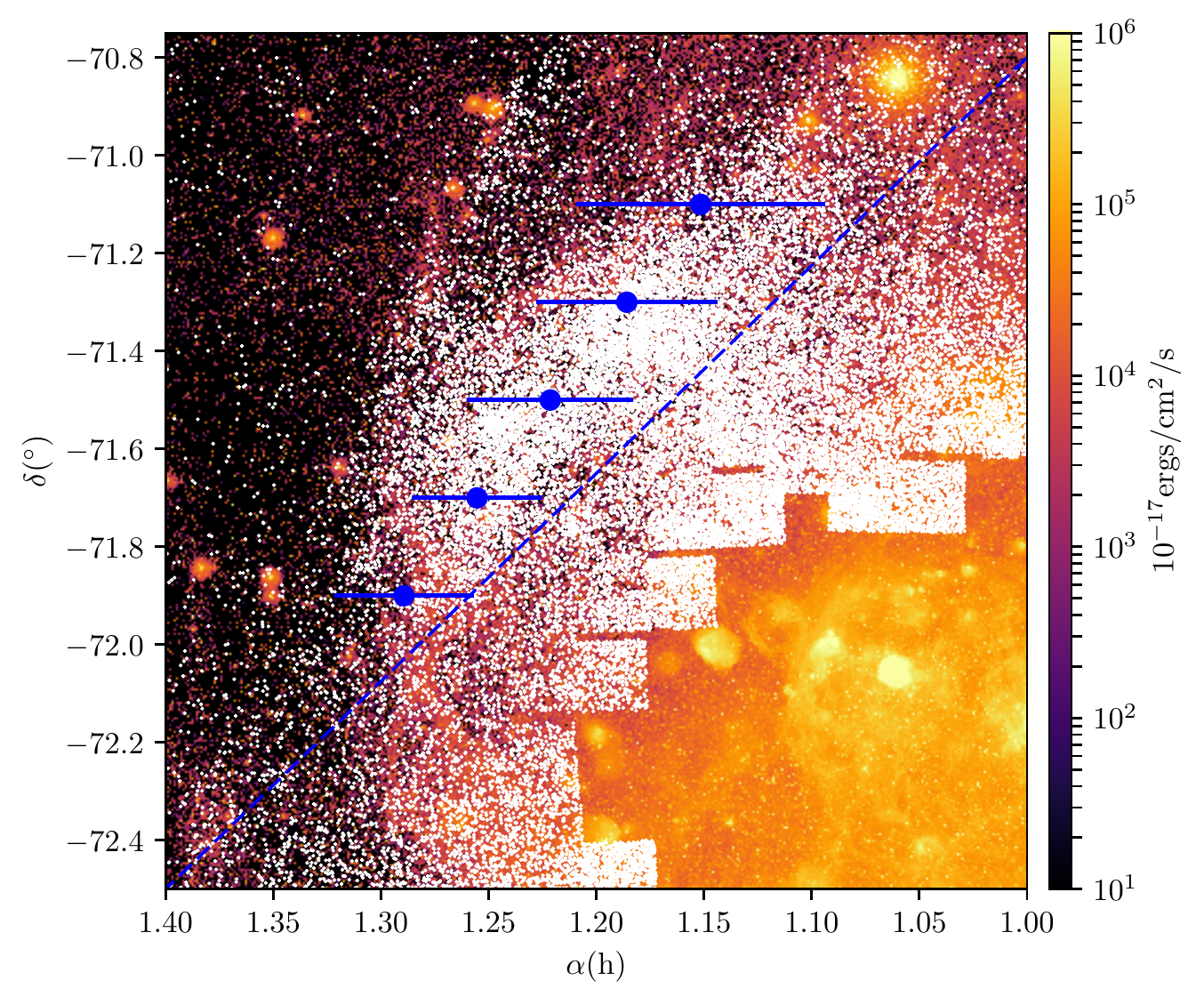} \includegraphics[width=0.5 \textwidth]{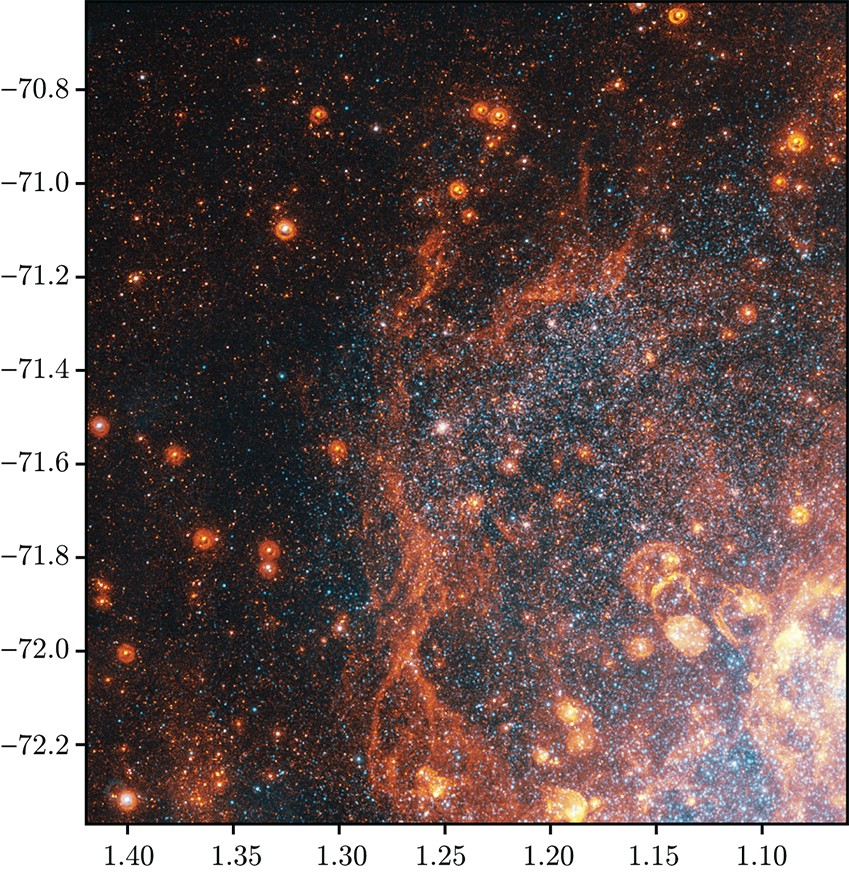}
  \caption{\emph{Left panel}: MCELS H$\alpha$ density map (red, following the color-bar density in the right side) compared to the young star density map plotted in Figure~4. Blue dots are the mean stellar position (and its dispersion as error bars) for stars located on the east of blue dashed line (on the right panel), grouped in 5 bins of declination ($-71 >\delta -72$). \emph{Right:} MCELS H$\alpha$ image zoomed in the region of the shell over-plotted to the GALEX image (see Fig. 5). It shows that a H$\alpha$ filament clearly overlaps the position of the spiral-arm-like feature (labelled {\it b} in Fig.4)}
    \label{fig-halfa}
\end{figure*}

As we do not have constraints to the distance of the H\,{\sc I} gas cloud in the Fig.~\ref{V198stars}, our few arguments are similarities between the velocities and shape of the H\,{\sc I} gas cloud and the shell-like feature of young stars in the Fig.~\ref{fig-maps}. Assuming that the `Z' shaped gas cloud was forming stars in a recent past (150-250 Myr ago), the velocities of the young stars and of the gas cloud could provide insights on what is the dominant process (tidal stripping or ram pressure) for decoupling the kinematics of the stars and the gas, at least in that region. However, if the gas cloud formed the shell-like feature, it is not clear what mechanism triggered such an intense star formation episode $\sim$200 Myr ago and why it is no longer acting to form stars. An evidence of the quiescence of the gas in forming stars is the lack of MS stars between the current position of the gas cloud and the position of the shell feature. Nevertheless, we must take into account the possibility of large distances in the line-of-sight between the gas feature and the young stars, being two completely separate substructures.


A comparison of the stellar populations map to the H$\alpha$ map (Fig. \ref{fig-halfa}) from the Magellanic Clouds Emission Line Survey (MCELS; Winkler, Rathore, \& Smith 1999) shows a system of filaments that roughly forms a shell with a radius of $0.75\degr$ centered at $\alpha_{J2000} = 01$:06, $\delta_{J2000} = -71$:45. This  interstellar material is reminiscent of supergiant shells similar to those catalogued in the Magellanic Clouds by Meaburn (1980). Such shells are believed to be driven by the collective action of winds from multiple OB associations and supernovae. When the H$\alpha$ images are compared to the H\,{\sc i} ATCA+Parkes observations, there appear to be faint (3-5$\sigma$  significance) H\,{\sc i} emission patches coincident with the H$\alpha$ filaments with velocities between $\sim$125-140.  We note that this putative shell does not coincide with those originally identified by Stanimirovic et al.\ (1999) using the same data set.

If this emission corresponds to the limb-brightened edge of a supergiant shell, the lack of detection of a coherent structure including envelopes expanding along the line-of-sight preclude a detailed kinematic comparison with the underlying stellar population.  The velocities of the faint H\,{\sc i} emission are consistent with the gas (and driving stellar population) being part of the SMC main body and not the material being drawn out by the interaction with the LMC. Based on the size, 750 pc (0.75\degr at 60 kpc) and assuming this interstellar structure arose from the collective stellar energy feedback from a population originating 150 Myr ago, we expect an expansion velocity of 5 km~s$^{-1}$ which is low but not unreasonable compared to other Magellanic supergiant shells (15 to 30 km~s$^{-1}$, Book et al.\ 2008).

\section{Discussion}

Our analysis of the resolved stellar populations of the elongated
over-density detected in our Canon 200 images reveals that it is the
brighter optical part of a more extended structure mainly traced by
blue, young stars distributed in an intricate and complex structure.
The structure shows a further, outer arc-like feature observed in the projected 
proximity of the Galactic globular cluster NGC 362 (see bottom panel of Figure~\ref{fig-maps}). The over-density also contains nine young star clusters with ages tightly clustered at 175 Myr (see Table~1).

We do not detect any counterpart of this over-density in the
distribution of the older stars in this area of the SMC. This dominant young age of our SMC over-density and the lack of an old stellar remnant in the stellar density map plotted in Fig.~\ref{fig-maps} (left panel) suggest that this feature is neither of tidal origin nor the stellar remnant of a tidally disrupted, lower-mass system as observed
in some other dwarf galaxies, e.g., in NGC 4449 (Mart\'\i nez-Delgado et al.\ 2012) or in Andromeda II (Amorisco et al.\ 2014). While recently a large number of
new, ultra-faint dwarf galaxies were discovered near the Magellanic
System (Drlica-Wagner et al.\ 2015, 2016), these systems are
exclusively devoid of gas and contain very old stellar populations.
The existence of a former, more gas-rich small dwarf irregular galaxy
is not excluded, but we would expect to also see a clear over-density in
the old stellar population map if such a system were to merge.  With the exception of
tidal dwarf galaxies, there are no nearby dwarf galaxies known that do
not contain old populations (Grebel \& Gallagher 2004).
Thus, the
nature of this over-density seems to be different from the one
discovered at $8\degr$ north of the SMC by Pieres et al.\ (2017),
which is mainly composed of intermediate-age stars and without
significant H\,{\sc i} gas. Although morphologically similar in appearance to the 
LMC substructure
described by Mackey et al. (2016), the young age of the SMC feature 
distinguishes it from that older and much larger LMC over-density.

Our stellar density maps of the shell region (Figure 4 and Figure 11) and the GALEX  data (Figure 5)
confirm with
higher resolution the asymmetric structure of young stars originally
found by Zaritsky et al.\ (2000) in the SMC outskirts, which contrasts
with the smooth distribution of the older stellar populations. This
lack of substructure in the older stellar populations led these
authors to conclude that the dominant physical mechanism in
determining the current appearance of the SMC must be recent star
formation possibly triggered by a hydrodynamic interaction of the SMC
with a second gas-rich object (e.g., a hot Milky Way halo or the outer
gaseous envelope of the LMC). The contrast between the small-scale spatial structures 
in the young
populations and the smooth distribution of the intermediate-age and
old populations are even starker in these new data. This finding
strengthens our argument (see also Zaritsky et al. 2000) that a purely
tidal origin for the feature, which would have affected stars of all
ages similarly, is unlikely. Instead, the origin of the feature is
most likely hydrodynamical in nature. Given the coherence of the
feature, the absence of an old stellar remnant, and the lack of evidence
for a strong shock in the H\,{\sc i} distribution, we disfavour models
involving a recent accretion event.

The elongated shape, estimated absolute magnitude (Sec.\ 3.2), and
position of the over-density in the outer region of the main body of
the SMC could also suggest a tidal dwarf galaxy as another possible
formation scenario for the shell-like structure. The stars in such 
objects are formed from gas stripped in past encounters (Elmegreen,
Kaufman \& Thomasson 1993) and/or consist of stars that originally
formed in the more massive galaxies participating in the interaction
(e.g., Duc 2012).  Key requirements for tidal dwarfs include that
they are made from recycled material, are gravitationally bound, and
have decoupled from their former parent (Duc 2012).  They may be able
to survive for at least 3 Gyr in spite of their lack of dark matter
(Ploeckinger et al.\ 2014).  In our case, the over-density appears to
be connected with the SMC and there is no evidence to suggest that it
is a separate entity. Hence we also discount the possibility of 
a tidal dwarf galaxy.

A conspicuous feature in the stellar density map is a
spiral-arm-like feature emanating from the over-density (labelled {\it b} in Figure 4). Because this stellar arm is only seen in
young stars, it is unlikely that it was produced via tidal
effects. Instead, the arc may consist of young stars formed as a result of
a low pitch angle spiral density wave in the outer gas.
Such features are common, but challenging to detect, in disks well beyond the optical radius (Ferguson et al. 1998; Herbert-Fort et al. 2012). The presence of a disk population would imply that the young stellar populations of the SMC must have some bulk rotation. 
However, initial {\it Gaia} DR2-based proper motions and rotation measurements of the SMC do not support significant rotation for the young SMC population (van der Marel \& Sahlmann 2016), suggesting that, if the density wave scenario is correct, then the observed offset is dominated by the progression of the pattern rather than motion of material through it. In addition, the hypothesis of a density wave also imply an expected offset between current (H$\alpha$) and past star formation. We find not detectable offset between the H$\alpha$ emission and GALEX sources (see Fig.~\ref{fig-halfa}) which, given the timescales these tracers are sensitive to (10$^7$ vs. 10$^8$ yrs), places an upper limit on the pattern speed. Within the new {\it Gaia} DR2 catalog, we find among the bright stars the same shell feature in the young stellar populations. Examining this subset of SMC stars reveals coherent motion among the stars but in an outward radial direction. Given the direction of motion, it continues to appear unlikely that the SMC possesses stellar rotation, at least rotation that lays primarily in the plane of the sky as with the neighboring LMC. Though, as the vast majority of these stars do not possess radial velocities, we are unable to fully assess if this motion may be due to rotation with a large inclination. A more holistic analysis of the SMC, including searching for other kinematic substructure, will be required to improve constraints on the degree of stellar rotation.


Our analysis of the stellar content of the shell (see Sec. 3.2) indicates that the main difference between the stellar population of the shell and the surrounding control fields is a recent period of enhanced star formation in the last $\simeq$ 1 Gyr, likely peaking at $\sim$ 150 Myr, as indicated by the clusters and Cepheids age distribution. The comparison of the CFs of the shell and the control fields also provides hints of a comparatively enhanced star formation rate in the total stellar content of this region at intermediate ages. If we compare these hints on the SFH of the shell and the control fields with the SFHs of the fields analyzed by No\"el  et al. (2009), it can be seen that the SFH of the shell field may present similarities with the SFHs of fields located in the Wing area of the SMC such as {\it qj0112}, {\it qj0111}, and {\it qj0116}, while the SFH of the two control fields may be alike to that of the remaining fields analyzed by No\"el et al. (2009), where SFH has been very low in the last $\simeq$ 1 Gyr. The qualitative hints on the SFH of the shell region obtained through the CF analysis are also consistent with the SFH for fields 1 and 4 (located in the bar and the wing area, respectively) analyzed by Cignoni et al. (2012), which present a strong enhancement from $\simeq$ 5 Gyr ago to the present time, and a strong peak at a very recent epoch, $<$ 200 Myr ago, and with the information on detailed SFH maps presented by Rubele et al. (2015), which indicate that the Wing is $<$ 200 Myr old, that metal-poor gas was injected $>$ 1 Gyr ago resulting in the formation of intermediate-age stars, and that the majority of the SMC mass resulted from a star formation episode 5 Gyr ago. The SFH of the Wing field, then, could be then similar to that of the SMC central body and Wing area, but characterized by an enhanced star formation rate in the last 200 Myr with respect to the past average star formation rate. However, since the CMD toward the shell region contains also a component of the field SMC population, it is possible that the intermediate-age populations that make up the CF peak at ($g-i)\sim$ 0 in Fig. 6 are part of that SMC underlying population. Therefore, there are two possible scenarios to explain the nature of the shell: i) a shell composed only of young stars resulting from a huge star forming region active $\sim$ 150 Myr ago; or ii) the shell is a region that had enhanced episodes of star formation at different epochs, with the most striking one happening $\sim$ 150 Myr ago.

It is, thus, interesting to consider the young age of our SMC
substructure in the context of the putative recent interaction with
the LMC about 100 to 300 Myr ago, which left its signature in the
young stellar populations in the Magellanic Bridge (e.g., Skowron et
al.\ 2014) and in the age distribution of young populous star clusters
in both Clouds (e.g., Glatt et al.\ 2010). Recently, Zivick et al. (2018) used mostly {\it Hubble Space Telescope} proper motion data 
to show that the LMC and SMC have had a head-on collision in the recent past, and it is able to constrain both the timescale and the impact parameter of this collision. 
Based on their measured proper motions and considering the allowed range of masses for the LMC and the Milky Way, they foud that in 97\% of all the considered cases,
the Clouds experienced a direct collision with each other 147 $\pm$ 33 Myr ago, with a mean impact parameter of 7.5 $\pm$ 2.5 kpc. There is also further evidence of this recent head-on collision  in the {\it Gaia} DR2 proper motions along the Bridge (Zivick et al. 2019). The age of our feature
falls into this age range as well. Moreover, the three-dimensional structure of young populations
traced by Cepheids (with ages of about 15--500 Myr) shows a highly asymmetric distribution throughout
the SMC: the distribution of classical Cepheids is elongated over 15--20 kpc (e.g., Scowcroft et al. 2016; Jacyszyn-Dobrzeniecka et al. 2016),
the NE region being younger, closer to the Sun than the SW region (e.g., Haschke et al. 2012; Subramanian et al. 2015; Ripepi et al. 2017). Haschke et al. (2012) speculate that the displacement and compression of the gas of the SMC through tidal and ram pressure effects caused by the 
interaction between the Magellanic Clouds (and with the Milky Way) may have locally enhanced 
star formation, creating some of the irregular, asymmetric features such as the over-density 
described in the current paper. Among others, Inno et al.(2015) and Jacyszyn-Dobrzeniecka et al.(2016) also propose that the concomitance of the extensive Cepheid formation in the LMC $\approx$140 Myr ago and of the younger episode of 
Cepheid formation in the SMC may be related to the interaction between the Clouds. Ripepi et al. (2017)
elaborate on this scenario and suggest that the relatively young Cepheids ($<$140 Myr) 
that dominate in the NE region have formed after the dynamical interaction that created 
the Bridge, from gas already shifted by the interaction.

Unfortunately, there is insufficient resolution in any existing
numerical simulations of the LMC-SMC interaction to see the fine
structure of our stellar density map in Fig. 4. Our results clearly
motivate more detailed studies of the internal structure and star
formation induced in the SMC by the LMC-SMC (and the MW)
interaction.  In particular, if the stellar arc is a spiral structure,
these data likely disfavor models where the SMC is originally modeled
as a non-rotating spheroid. This result further highlights the
discrepant kinematics and spatial distribution of the SMC younger and
older stellar populations, the origin of which is currently unknown.

\begin{acknowledgements}

We thank David Hogg for his help with the astrometry solution of the image used in this work. DMD thanks Prof. Ken Freeman for a discussion about the detection of the SMC shell in the photographic plates of the Clouds taken in the 1950s. DMD also thanks the hospitality and fruitful discussion about this work with the European Southern Observatory Garching headquarters staff during his stay as part of the ESO visitor program in September 2017. DMD, EKG, BL and LI acknowledge support by Sonderforschungsbereich (SFB) 881 ``The Milky Way System'' of the German Research Foundation (DFG), particularly through sub-projects A2, A3 and A5. DMD acknowledges support from the Spanish MINECO grant AYA2016-81065-C2-2. This project used data obtained with the Dark Energy Camera (DECam), which was constructed by the Dark Energy Survey (DES) collaboration. M-RC and CB acknowledge support from the European Research Council (ERC) under the European Union$'$s Horizon 2020 research and innovation program (grant agreement No. 682115). BCC acknowledges the support of the Australian Research Council through Discovery project DP150100862. T.d.B. acknowledges support from the European Research Council (ERC StG-335936). This work has been supported by the Spanish Ministry of Economy and Competitiveness (MINECO) under grant AYA2014-56795-P. MS acknowledges support from  the ADAP grant NNX14AF81G. R.R.M. acknowledges partial support from project BASAL AFB-$170002$ as well as FONDECYT project N$^{\circ}1170364$.Y.C. acknowledges support from NSF grant AST 1655677.

Based on observations at Cerro Tololo Inter-American Observatory, National Optical Astronomy Observatory which is operated by the Association of Universities for Research in Astronomy (AURA) under a cooperative agreement with the National Science Foundation.
This project used data obtained with the Dark Energy Camera (DECam), which was constructed by the Dark Energy Survey (DES) collaboration. Funding for the DES Projects has been provided by 
the U.S. Department of Energy, 
the U.S. National Science Foundation, 
the Ministry of Science and Education of Spain, 
the Science and Technology Facilities Council of the United Kingdom, 
the Higher Education Funding Council for England, 
the National Center for Supercomputing Applications at the University of Illinois at Urbana-Champaign, 
the Kavli Institute of Cosmological Physics at the University of Chicago, 
the Center for Cosmology and Astro-Particle Physics at the Ohio State University, 
the Mitchell Institute for Fundamental Physics and Astronomy at Texas A\&M University, 
Financiadora de Estudos e Projetos, Funda{\c c}{\~a}o Carlos Chagas Filho de Amparo {\`a} Pesquisa do Estado do Rio de Janeiro, 
Conselho Nacional de Desenvolvimento Cient{\'i}fico e Tecnol{\'o}gico and the Minist{\'e}rio da Ci{\^e}ncia, Tecnologia e Inovac{\~a}o, 
the Deutsche Forschungsgemeinschaft, 
and the Collaborating Institutions in the Dark Energy Survey. This work has made use of data from the European Space Agency (ESA) mission {\it Gaia} (\url{<https://www.cosmos.esa.int/gaia}), processed by the {\it Gaia} Data Processing and Analysis Consortium (DPAC, \url{<https://www.cosmos.esa.int/web/gaia/dpac/consortium}). Funding for the DPC has been provided by national institutions, in particular the institutions participating in the {\it Gaia} Multilateral Agreement.
\end{acknowledgements}



\begin{thebibliography}{99}

\bibitem[Abbott et al.(2018)]{2018ApJS..239...18A} Abbott, T.~M.~C., Abdalla, F.~B., Allam, S., et al.\ 2018, \apjs, 239, 18 
\bibitem[Albers et al.(1987)]{1987A&A...182L...8A} Albers, H., MacGillivray, H.~T., Beard, S.~M., \& Chromey, F.~R.\ 1987, \aap, 182, L8 
\bibitem[Alcaino et al.(2003)]{2003A&A...400..917A} Alcaino, G., Alvarado, F., Borissova, J., \& Kurtev, R.\ 2003, \aap, 400, 917 
\bibitem[Amorisco et al.(2014)]{2014Natur.507..335A} Amorisco, N.~C., Evans, N.~W., \& van de Ven, G.\ 2014, \nat, 507, 335 
\bibitem[Anderson et al.(2016)]{Ander2016} Anderson, R. I., Saio, H., Ekstr\"om, S., Georgy, C., Meynet, G, 2016,
      A\&A 591, A8
\bibitem[Bekki \& Chiba(2008)]{2008ApJ...679L..89B} Bekki, K., \& Chiba, M.\ 2008, \apjl, 679, L89 
\bibitem[Bekki(2011)]{2011MNRAS.416.2359B} Bekki, K.\ 2011, \mnras, 416, 2359 
\bibitem[Belokurov \& Koposov(2016)]{2016MNRAS.456..602B} Belokurov, V., \& Koposov, S.~E.\ 2016, \mnras, 456, 602 
\bibitem[Belokurov et al.(2017)]{2017MNRAS.466.4711B} Belokurov, V., Erkal, D., Deason, A.~J., et al.\ 2017, \mnras, 466, 4711 
\bibitem[Bertin \& Arnouts(1996)]{1996A&AS..117..393B} Bertin, E., \& Arnouts, S.\ 1996, \aaps, 117, 393 
\bibitem[Besla et al.(2007)]{2007ApJ...668..949B} Besla, G., Kallivayalil, N., Hernquist, L., et al.\ 2007, \apj, 668, 949 
\bibitem[Besla et al.(2010)]{2010ApJ...721L..97B} Besla, G., Kallivayalil, N., Hernquist, L., et al.\ 2010, \apjl, 721, L97 
\bibitem[Besla et al.(2016)]{2016ApJ...825...20B} Besla, G., Mart{\'{\i}}nez-Delgado, D., van der Marel, R.~P., et al.\ 2016, \apj, 825, 20 
\bibitem[Bica et al.(2008)]{Bica2008} Bica, E., Bonatto, C., Dutra, C.~M., \& Santos, J.~F.~C.\ 2008, \mnras, 389, 678 
\bibitem[Bono et al.(2005)]{Bono2005} Bono, G., Marconi, M., Cassisi, S., Caputo, F., Gieren, W., Pietrzy\'nski, G., 2005
      ApJ 621, 966
\bibitem[Book et al.(2008)]{2008ApJS..175..165B} Book, L.~G., Chu, Y.-H., \& Gruendl, R.~A.\ 2008, \apjs, 175, 165 
\bibitem[Boylan-Kolchin et al.(2011)]{2011MNRAS.414.1560B} Boylan-Kolchin, M., Besla, G., \& Hernquist, L.\ 2011, \mnras, 414, 1560 
\bibitem[Bruck(1980)]{1980A&A....87...92B} Bruck, M.~T.\ 1980, \aap, 87, 92  
\bibitem[Brueck \& Marsoglu(1978)]{1978A&A....68..193B} Brueck, M.~T., \& Marsoglu, A.\ 1978, \aap, 68, 193 
\bibitem[Busha et al.(2011)]{2011ApJ...743...40B} Busha, M.~T., Marshall, P.~J., Wechsler, R.~H., Klypin, A., \& Primack, J.\ 2011, \apj, 743, 40 
\bibitem[Caldwell \& Coulson(1986)]{1986MNRAS.218..223C} Caldwell, J.~A.~R., \& Coulson, I.~M.\ 1986, \mnras, 218, 223
\bibitem[Carballo-Bello(2019)]{2019MNRAS.486.1667C} Carballo-Bello, J.~A.\ 2019, \mnras, 486, 1667 
\bibitem[Carrera et al.(2008)]{2008AJ....136.1039C} Carrera, R., Gallart, C., Aparicio, A., et al.\ 2008, \aj, 136, 1039 
\bibitem[Carrera et al.(2017)]{2017MNRAS.471.4571C} Carrera, R., Conn, B.~C., No{\"e}l, N.~E.~D., Read, J.~I., \& L{\'o}pez S{\'a}nchez, {\'A}.~R.\ 2017, \mnras, 471, 4571 
\bibitem[Casetti-Dinescu et al.(2014)]{2014ApJ...784L..37C} Casetti-Dinescu, D.~I., Moni Bidin, C., Girard, T.~M., et al.\ 2014, \apjl, 784, L37 
\bibitem[Choi et al.(2018a)]{2018ApJ...866...90C} Choi, Y., Nidever, D.~L., Olsen, K., et al.\ 2018, \apj, 866, 90 
\bibitem[Choi et al.(2018b)]{2018ApJ...869..125C} Choi, Y., Nidever, D.~L., Olsen, K., et al.\ 2018, \apj, 869, 125 
\bibitem[Cignoni et al.(2012)]{2012ApJ...754..130C} Cignoni, M., Cole, A.~A., Tosi, M., et al.\ 2012, \apj, 754, 130 
\bibitem[Cignoni et al.(2013)]{2013ApJ...775...83C} Cignoni, M., Cole, A.~A., Tosi, M., et al.\ 2013, \apj, 775, 83 
\bibitem[Cioni et al.(2000)]{2000A&A...358L...9C} Cioni, M.-R.~L., Habing, H.~J., \& Israel, F.~P.\ 2000, \aap, 358, L9 
\bibitem[Crowl et al.(2001)]{2001AJ....122..220C} Crowl, H.~H., Sarajedini, A., Piatti, A.~E., et al.\ 2001, \aj, 122, 220 
\bibitem[de Grijs \& Bono(2015)]{2015AJ....149..179D} de Grijs, R., \& Bono, G.\ 2015, \aj, 149, 179 
\bibitem[de Vaucouleurs \& Freeman(1972)]{1972VA.....14..163D} de Vaucouleurs, G., \& Freeman, K.~C.\ 1972, Vistas in Astronomy, 14, 163 
\bibitem[DOnghia et al. 2009]{2009Nature} D\' Onghia, E., Besla, G., Cox, T. J., Hernquist, L., 2009, Nature, 460, 605
\bibitem[D'Onghia \& Fox(2016)]{2016ARA&A..54..363D} D'Onghia, E., \& Fox, A.~J.\ 2016, \araa, 54, 363 
\bibitem[Drlica-Wagner et al.(2015)]{2015ApJ...813..109D} Drlica-Wagner, A., Bechtol, K., Rykoff, E.~S., et al.\ 2015, \apj, 813, 109 
\bibitem[Drlica-Wagner et al.(2016)]{2016ApJ...833L...5D} Drlica-Wagner, A., Bechtol, K., Allam, S., et al.\ 2016, \apjl, 833, L5 
\bibitem[Duc(2012)]{2012ASSP...28..305D} Duc, P.-A.\ 2012, Astrophysics and Space Science Proceedings, 28, 305 
\bibitem[Ederoclite \& Cepa(2010)]{2010ASPC..434..253E} Ederoclite, A., \& Cepa, J.\ 2010, Astronomical Data Analysis Software and Systems XIX, 434, 253 
\bibitem[Elmegreen et al.(1993)]{1993ApJ...412...90E} Elmegreen, B.~G., Kaufman, M., \& Thomasson, M.\ 1993, \apj, 412, 90 
\bibitem[Evans \& Howarth(2008)]{2008MNRAS.386..826E} Evans, C.~J., \& Howarth, I.~D.\ 2008, \mnras, 386, 826 
\bibitem[Fox et al.(2014)]{2014ApJ...787..147F} Fox, A.~J., Wakker, B.~P., Barger, K.~A., et al.\ 2014, \apj, 787, 147 
\bibitem[Gaia Collaboration et al.(2018)]{2018A&A...616A...1G} Gaia Collaboration, Brown, A.~G.~A., Vallenari, A., et al.\ 2018, \aap, 616, A1 
\bibitem[Gaia Collaboration et al.(2018)]{2018A&A...616A..12G} Gaia Collaboration, Helmi, A., van Leeuwen, F., et al.\ 2018, \aap, 616, A12 
\bibitem[Gallart et al.(2005)]{2005ARA&A..43..387G} Gallart, C., Zoccali, M., \& Aparicio, A.\ 2005, \araa, 43, 387 
\bibitem[Glatt et al.(2010)]{2010A&A...517A..50G} Glatt, K., Grebel, E.~K., \& Koch, A.\ 2010, \aap, 517, A50 
\bibitem[Gonz{\'a}lez et al.(2013)]{2013ApJ...770...96G} Gonz{\'a}lez, R.~E., Kravtsov, A.~V., \& Gnedin, N.~Y.\ 2013, \apj, 770, 96 
\bibitem[Grebel \& Gallagher(2004)]{2004ApJ...610L..89G} Grebel, E.~K., \& Gallagher, J.~S., III 2004, \apjl, 610, L89 
\bibitem[Harris \& Zaritsky(2002)]{2002ASPC..285..313H} Harris, J., \& Zaritsky, D.\ 2002, Modes of Star Formation and the Origin of Field Populations, 285, 313 
\bibitem[Haschke et al.(2012)]{Has2012b} Haschke, R., Grebel, E. K., Duffau, S., 2012, AJ 144, 107   
\bibitem[Herbert-Fort et al.(2012)]{2012ApJ...754..110H} Herbert-Fort, S., Zaritsky, D., Moustakas, J., et al.\ 2012, \apj, 754, 110 
\bibitem[Inno et al.(2015)]{Inno2015a} Inno, L., Bono, G., Romaniello, M., Matsunaga, N., Pietrinferni, A., Genovali, K., Lemasle, B., Marconi, M., Primas, F., 2015 ASPC 491, 265
\bibitem[Irwin et al.(1985)]{1985Natur.318..160I} Irwin, M.~J., Kunkel, W.~E., \& Demers, S.\ 1985, \nat, 318, 160 
\bibitem[Jacobs et al.(2009)]{2009AJ....138..332J} Jacobs, B.~A., Rizzi, L., Tully, R.~B., et al.\ 2009, \aj, 138, 332 
\bibitem[Jacyszyn-Dobrzeniecka et al.(2016)]{Jac2016} Jacyszyn-Dobrzeniecka, A. M., Skowron, D. M., Mr\'oz, P., Skowron, J., Soszy\'nski, I., Udalski, A., Pietrukowicz, P., Kozlowski, S., Wyrzykowski, L., Poleski, R., Pawlak, M., Szyma\'nski, M. K., Ulaczyk, K., 2016,
      AcA 66, 149
\bibitem[Jacyszyn-Dobrzeniecka et al.(2017)]{2017AcA....67....1J} Jacyszyn-Dobrzeniecka, A.~M., Skowron, D.~M., Mr{\'o}z, P., et al.\ 2017, \actaa, 67, 1 
\bibitem[Jacyszyn-Dobrzeniecka et al.(2019)]{2019arXiv190408220J} Jacyszyn-Dobrzeniecka, A.~M., Soszy{\'n}ski, I., Udalski, A., et al.\ 2019, arXiv:1904.08220       
\bibitem[Kalberla \& Haud(2015)]{2015A&A...578A..78K} Kalberla, P.~M.~W., \& Haud, U.\ 2015, \aap, 578, A78 
\bibitem[Kallivayalil et al.(2006)]{2006ApJ...652.1213K} Kallivayalil, N., van der Marel, R.~P., \& Alcock, C.\ 2006, \apj, 652, 1213 
\bibitem[Kallivayalil et al.(2006)]{2006ApJ...638..772K} Kallivayalil, N., van der Marel, R.~P., Alcock, C., et al.\ 2006, \apj, 638, 772 
\bibitem[Kallivayalil et al.(2013)]{2013ApJ...764..161K} Kallivayalil, N., van der Marel, R.~P., Besla, G., Anderson, J., \& Alcock, C.\ 2013, \apj, 764, 161 
\bibitem[Kroupa et al.(1993)]{1993MNRAS.262..545K} Kroupa, P., Tout, C.~A., \& Gilmore, G.\ 1993, \mnras, 262, 545 
\bibitem[Lang et al.(2010)]{2010AJ....139.1782L} Lang, D., Hogg, D.~W., Mierle, K., Blanton, M., \& Roweis, S.\ 2010, \aj, 139, 1782 
\bibitem[Lemasle et al.(2017)]{2017A&A...608A..85L} Lemasle, B., Groenewegen, M.~A.~T., Grebel, E.~K., et al.\ 2017, \aap, 608, A85 
\bibitem[Mackey et al.(2016)]{2016MNRAS.459..239M} Mackey, A.~D., Koposov, S.~E., Erkal, D., et al.\ 2016, \mnras, 459, 239 
\bibitem[Mackey et al.(2018)]{2018ApJ...858L..21M} Mackey, D., Koposov, S., Da Costa, G., et al.\ 2018, \apjl, 858, L21 
\bibitem[Mart{\'{\i}}nez-Delgado et al.(2012)]{2012ApJ...748L..24M} Mart{\'{\i}}nez-Delgado, D., Romanowsky, A.~J., Gabany, R.~J., et al.\ 2012, \apjl, 748, L24 
\bibitem[Mart{\'{\i}}nez-Delgado et al.(2015)]{2015AJ....150..116M} Mart{\'{\i}}nez-Delgado, D., D'Onghia, E., Chonis, T.~S., et al.\ 2015, \aj, 150, 116 
\bibitem[Mathewson et al.(1974)]{1974ApJ...190..291M} Mathewson, D.~S., Cleary, M.~N., \& Murray, J.~D.\ 1974, \apj, 190, 291 
\bibitem[Mathewson et al.(1988)]{1988ApJ...333..617M} Mathewson, D.~S., Ford, V.~L., \& Visvanathan, N.\ 1988, \apj, 333, 617 
\bibitem[Meaburn(1980)]{1980MNRAS.192..365M} Meaburn, J.\ 1980, \mnras, 192, 365 
\bibitem[Muraveva et al.(2018)]{2018MNRAS.473.3131M} Muraveva, T., Subramanian, S., Clementini, G., et al.\ 2018, \mnras, 473, 3131 
\bibitem[Nidever et al.(2010)]{2010ApJ...723.1618N} Nidever, D.~L., Majewski, S.~R., Butler Burton, W., \& Nigra, L.\ 2010, \apj, 723, 1618 
\bibitem[Nidever et al.(2011)]{2011ApJ...733L..10N} Nidever, D.~L., Majewski, S.~R., Mu{\~n}oz, R.~R., et al.\ 2011, \apjl, 733, L10 
\bibitem[Nidever et al.(2013)]{2013ApJ...779..145N} Nidever, D.~L., Monachesi, A., Bell, E.~F., et al.\ 2013, \apj, 779, 145 
\bibitem[Nidever et al.(2017)]{2017AJ....154..199N} Nidever, D.~L., Olsen, K., Walker, A.~R., et al.\ 2017, \aj, 154, 199 
\bibitem[No{\"e}l \& Gallart(2007)]{2007ApJ...665L..23N} No{\"e}l, N.~E.~D., \& Gallart, C.\ 2007, \apjl, 665, L23 
\bibitem[No{\"e}l et al.(2009)]{2009ApJ...705.1260N} No{\"e}l, N.~E.~D., Aparicio, A., Gallart, C., et al.\ 2009, \apj, 705, 1260 
\bibitem[No{\"e}l et al.(2013)]{2013ApJ...768..109N} No{\"e}l, N.~E.~D., Conn, B.~C., Carrera, R., et al.\ 2013, \apj, 768, 109 
\bibitem[No{\"e}l et al.(2015)]{2015MNRAS.452.4222N} No{\"e}l, N.~E.~D., Conn, B.~C., Read, J.~I., et al.\ 2015, \mnras, 452, 4222

\bibitem[Pasquali et al. 2005]{2005AJ....129..148P} Pasquali, A., Larsen, S., Ferreras, I., et al.\ 2005, \aj, 129, 148 
\bibitem[Piatek et al.(2008)]{2008AJ....135.1024P} Piatek, S., Pryor, C., \& Olszewski, E.~W.\ 2008, \aj, 135, 1024 
\bibitem[Piatti(2014)]{2014MNRAS.445.2302P} Piatti, A.~E.\ 2014, \mnras, 445, 2302 
\bibitem[Pieres et al.(2017)]{2017MNRAS.468.1349P} Pieres, A., Santiago, B.~X., Drlica-Wagner, A., et al.\ 2017, \mnras, 468, 1349 
\bibitem[Pietrinferni et al.(2004)]{2004ApJ...612..168P} Pietrinferni, A., Cassisi, S., Salaris, M., \& Castelli, F.\ 2004, \apj, 612, 168 
\bibitem[Putman et al.(1998)]{1998Natur.394..752P} Putman, M.~E., Gibson, B.~K., Staveley-Smith, L., et al.\ 1998, \nat, 394, 752 
\bibitem[Ripepi et al.(2017)]{Ripepi2017} Ripepi, V., Cioni, M.-R. L., Moretti, M. I., Marconi, M., Bekki, K., Clementini, G., de Grijs, R., Emerson, J., Groenewegen, M. A. T., Ivanov, V. D., Molinaro, R., Muraveva, T., Oliveira, J. M., Piatti, A. E., Subramanian, S., van Loon, J. Th.,
 2017, MNRAS 472, 808	
 \bibitem[Rodr{\'{\i}}guez-Puebla et al.(2013)]{2013ApJ...773..172R} Rodr{\'{\i}}guez-Puebla, A., Avila-Reese, V., \& Drory, N.\ 2013, \apj, 773, 172 \bibitem[Romaniello et al.(2008)]{Roma2008} Romaniello, M., Primas, F., Mottini, M., Pedicelli, S., Lemasle, B.,  Bono, G., Fran\c cois, P., Groenewegen, M. A. T., Laney, C. D., 2008,  A\&A 488, 731
\bibitem[Rubele et al.(2015)]{2015MNRAS.449..639R} Rubele, S., Girardi, L., Kerber, L., et al.\ 2015, \mnras, 449, 639 
\bibitem[Rubele et al.(2018)]{2018MNRAS.478.5017R} Rubele, S., Pastorelli, G., Girardi, L., et al.\ 2018, \mnras, 478, 5017 
\bibitem[Russell \& Dopita(1992)]{1992ApJ...384..508R} Russell, S.~C., \& Dopita, M.~A.\ 1992, \apj, 384, 508 
\bibitem[Schlegel et al.(1998)]{1998ApJ...500..525S} Schlegel, D.~J., Finkbeiner, D.~P., \& Davis, M.\ 1998, \apj, 500, 525 
\bibitem[Scowcroft et al.(2016)]{Sco2016} Scowcroft, V., Freedman, W., L., Madore, B. F., Monson, A. J., Persson, S. E., Rich, J., Seibert, M., Rigby, J. R., 2016 ApJ 816, 49S
\bibitem[Schlafly \& Finkbeiner(2011)]{2011ApJ...737..103S} Schlafly, E.~F., \& Finkbeiner, D.~P.\ 2011, \apj, 737, 103 
\bibitem[Skowron et al.(2014)]{2014ApJ...795..108S} Skowron, D.~M., Jacyszyn, A.~M., Udalski, A., et al.\ 2014, \apj, 795, 108 
\bibitem[Soszy\'nski et al.(2016)]{Sos2016} Soszy\'nski, I., Udalski, A., Szyma\'nski, M. K., Skowron, D., Pietrzy\'nski, G., Poleski, R., Pietrukowicz, P.,  Skowron, J., Mr\'oz, P., Kozlowski, S., Wyrzykowski, L., Ulaczyk, K.,  Pawlak, M., 2016,
     AcA 65, 297
\bibitem[Stanimirovic et al.(1999)]{1999MNRAS.302..417S} Stanimirovic, S., Staveley-Smith, L., Dickey, J.~M., Sault, R.~J., \& Snowden, S.~L.\ 1999, \mnras, 302, 417 
\bibitem[Stanimirovi{\'c} et al.(2004)]{2004ApJ...604..176S} Stanimirovi{\'c}, S., Staveley-Smith, L., \& Jones, P.~A.\ 2004, \apj, 604, 176 
\bibitem[Stetson(1987)]{1987PASP...99..191S} Stetson, P.B. 1987, \pasp, 99, 191
\bibitem[Subramanian \& Subramaniam(2012)]{2012ApJ...744..128S} Subramanian, S., \& Subramaniam, A.\ 2012, \apj, 744, 128 
\bibitem[Subramanian et al.(2015)]{Subra2015} Subramanian, S., Subramaniam, A., 2015,
      A\&A 573, A135
\bibitem[Subramanian et al.(2017)]{2017MNRAS.467.2980S} Subramanian, S., Rubele, S., Sun, N.-C., et al.\ 2017, \mnras, 467, 2980 
\bibitem[Udalski et al.(2015)]{Udal2015} Udalski, A., Szymanski, M. K., Szymanski, G., 2015,
      AcA 65, 1
\bibitem[van der Marel et al.(2002)]{2002AJ....124.2639V} van der Marel, R.~P., Alves, D.~R., Hardy, E., \& Suntzeff, N.~B.\ 2002, \aj, 124, 2639 
\bibitem[van der Marel \& Sahlmann(2016)]{2016ApJ...832L..23V} van der Marel, R.~P., \& Sahlmann, J.\ 2016, \apjl, 832, L23 
\bibitem[Weisz et al.(2013)]{2013MNRAS.431..364W} Weisz, D.~R., Dolphin, A.~E., Skillman, E.~D., et al.\ 2013, \mnras, 431, 364 
\bibitem[Winkler et al.(1999)]{1999IAUS..190...97W} Winkler, P.~F., Rathore, Y., \& Smith, R.~C.\ 1999, New Views of the Magellanic Clouds, 190, 97 
\bibitem[Zaritsky et al.(2000)]{2000ApJ...534L..53Z} Zaritsky, D., Harris, J., Grebel, E.~K., \& Thompson, I.~B.\ 2000, \apjl, 534, L53 
\bibitem[Zivick et al.(2018)]{2018ApJ...864...55Z} Zivick, P., Kallivayalil, N., van der Marel, R.~P., et al.\ 2018, \apj, 864, 55 
\bibitem[Zivick et al.(2019)]{2019ApJ...874...78Z} Zivick, P., Kallivayalil, N., Besla, G., et al.\ 2019, \apj, 874, 78 
\end{thebibliography}
\end{document}